\documentclass[ALICE,manyauthors]{cernphprep}
\usepackage[comma,square,numbers,sort&compress]{natbib}
\usepackage{hyperref}
\usepackage{lineno}
\usepackage{xspace}
\usepackage{comment}
\usepackage{rotating}
\usepackage[utf8]{inputenc}
\usepackage{amsmath}
\usepackage[T1]{fontenc}

\begin{document}
%

\newcommand{\pp}           {pp\xspace}
\newcommand{\ppbar}        {\mbox{$\mathrm {p\overline{p}}$}\xspace}
\newcommand{\XeXe}         {\mbox{Xe--Xe}\xspace}
\newcommand{\PbPb}         {\mbox{Pb--Pb}\xspace}
\newcommand{\pA}           {\mbox{pA}\xspace}
\newcommand{\pPb}          {\mbox{p--Pb}\xspace}
\newcommand{\AuAu}         {\mbox{Au--Au}\xspace}
\newcommand{\dAu}          {\mbox{d--Au}\xspace}

\newcommand{\s}            {\ensuremath{\sqrt{s}}\xspace}
\newcommand{\snn}          {\ensuremath{\sqrt{s_{\mathrm{NN}}}}\xspace}
\newcommand{\pt}           {\ensuremath{p_{\rm T}}\xspace}
\newcommand{\meanpt}       {$\langle p_{\mathrm{T}}\rangle$\xspace}
\newcommand{\ycms}         {\ensuremath{y_{\rm CMS}}\xspace}
\newcommand{\ylab}         {\ensuremath{y_{\rm lab}}\xspace}
\newcommand{\etarange}[1]  {\mbox{$\left | \eta \right |~<~#1$}}
\newcommand{\yrange}[1]    {\mbox{$\left | y \right |~<~#1$}}
\newcommand{\dndy}         {\ensuremath{\mathrm{d}N_\mathrm{ch}/\mathrm{d}y}\xspace}
\newcommand{\dndeta}       {\ensuremath{\mathrm{d}N_\mathrm{ch}/\mathrm{d}\eta}\xspace}
\newcommand{\avdndeta}     {\ensuremath{\langle\dndeta\rangle}\xspace}
\newcommand{\dNdy}         {\ensuremath{\mathrm{d}N_\mathrm{ch}/\mathrm{d}y}\xspace}
\newcommand{\Npart}        {\ensuremath{N_\mathrm{part}}\xspace}
\newcommand{\Ncoll}        {\ensuremath{N_\mathrm{coll}}\xspace}
\newcommand{\dEdx}         {\ensuremath{\textrm{d}E/\textrm{d}x}\xspace}
\newcommand{\RpPb}         {\ensuremath{R_{\rm pPb}}\xspace}

\newcommand{\nineH}        {$\sqrt{s}~=~0.9$~Te\kern-.1emV\xspace}
\newcommand{\seven}        {$\sqrt{s}~=~7$~Te\kern-.1emV\xspace}
\newcommand{\twoH}         {$\sqrt{s}~=~0.2$~Te\kern-.1emV\xspace}
\newcommand{\twosevensix}  {$\sqrt{s}~=~2.76$~Te\kern-.1emV\xspace}
\newcommand{\five}         {$\sqrt{s}~=~5.02$~Te\kern-.1emV\xspace}
\newcommand{\twosevensixnn}{$\sqrt{s_{\mathrm{NN}}}~=~2.76$~Te\kern-.1emV\xspace}
\newcommand{\fivenn}       {$\sqrt{s_{\mathrm{NN}}}~=~5.02$~Te\kern-.1emV\xspace}
\newcommand{\LT}           {L{\'e}vy-Tsallis\xspace}
\newcommand{\GeVc}         {Ge\kern-.1emV/$c$\xspace}
\newcommand{\MeVc}         {Me\kern-.1emV/$c$\xspace}
\newcommand{\TeV}          {Te\kern-.1emV\xspace}
\newcommand{\GeV}          {Ge\kern-.1emV\xspace}
\newcommand{\MeV}          {Me\kern-.1emV\xspace}
\newcommand{\GeVmass}      {Ge\kern-.2emV/$c^2$\xspace}
\newcommand{\MeVmass}      {Me\kern-.2emV/$c^2$\xspace}
\newcommand{\lumi}         {\ensuremath{\mathcal{L}}\xspace}

\newcommand{\ITS}          {\rm{ITS}\xspace}
\newcommand{\TOF}          {\rm{TOF}\xspace}
\newcommand{\ZDC}          {\rm{ZDC}\xspace}
\newcommand{\ZDCs}         {\rm{ZDCs}\xspace}
\newcommand{\ZNA}          {\rm{ZNA}\xspace}
\newcommand{\ZNC}          {\rm{ZNC}\xspace}
\newcommand{\SPD}          {\rm{SPD}\xspace}
\newcommand{\SDD}          {\rm{SDD}\xspace}
\newcommand{\SSD}          {\rm{SSD}\xspace}
\newcommand{\TPC}          {\rm{TPC}\xspace}
\newcommand{\TRD}          {\rm{TRD}\xspace}
\newcommand{\VZERO}        {\rm{V0}\xspace}
\newcommand{\VZEROA}       {\rm{V0A}\xspace}
\newcommand{\VZEROC}       {\rm{V0C}\xspace}
\newcommand{\Vdecay} 	   {\ensuremath{V^{0}}\xspace}

\newcommand{\ee}           {\ensuremath{e^{+}e^{-}}\xspace} 
\newcommand{\pip}          {\ensuremath{\pi^{+}}\xspace}
\newcommand{\pim}          {\ensuremath{\pi^{-}}\xspace}
\newcommand{\kap}          {\ensuremath{\rm{K}^{+}}\xspace}
\newcommand{\kam}          {\ensuremath{\rm{K}^{-}}\xspace}
\newcommand{\pbar}         {\ensuremath{\rm\overline{p}}\xspace}
\newcommand{\kzero}        {\ensuremath{{\rm K}^{0}_{\rm{S}}}\xspace}
\newcommand{\lmb}          {\ensuremath{\Lambda}\xspace}
\newcommand{\almb}         {\ensuremath{\overline{\Lambda}}\xspace}
\newcommand{\Om}           {\ensuremath{\Omega^-}\xspace}
\newcommand{\Mo}           {\ensuremath{\overline{\Omega}^+}\xspace}
\newcommand{\X}            {\ensuremath{\Xi^-}\xspace}
\newcommand{\Ix}           {\ensuremath{\overline{\Xi}^+}\xspace}
\newcommand{\Xis}          {\ensuremath{\Xi^{\pm}}\xspace}
\newcommand{\Oms}          {\ensuremath{\Omega^{\pm}}\xspace}
\newcommand{\degree}       {\ensuremath{^{\rm o}}\xspace}
\newcommand{\mm}           {\ensuremath{\mu^{+}\mu^{-}}\xspace}

\newcommand{\Dzero}     {\ensuremath{\rm D}^{0}\xspace}
\newcommand{\Dplus}     {\ensuremath{\rm D}^{+}\xspace}
\newcommand{\Dstar}     {\ensuremath{\rm D}^{*+}\xspace}
\newcommand{\Dstarwide} {\ensuremath{\rm D}^{*}(2010)^{+}\xspace}
\newcommand{\Dphi}      {\Delta\varphi}
\newcommand{\Deta}      {\Delta\eta}
\newcommand{\ptD}       {\ensuremath{p_{\rm T}}^{\rm D}\xspace}
\newcommand{\ptass}     {\ensuremath{p_{\rm T}}^{\rm assoc}\xspace}

\mathchardef\mhyphen="2D

\begin{titlepage}
\PHyear{2021}       
\PHnumber{184}      
\PHdate{10 September}  

\title{Investigating charm production and fragmentation via azimuthal correlations of prompt D mesons with charged particles \linebreak in pp collisions at $\mathbf{\sqrt{\it s} = 13}$ \TeV}
\ShortTitle{D meson-charged particle azimuthal correlations at 13 TeV}   

\Collaboration{ALICE Collaboration\thanks{See Appendix~\ref{app:collab} for the list of collaboration members}}
\ShortAuthor{ALICE Collaboration} 

\begin{abstract}
Angular correlations of heavy-flavour and charged particles in high-energy proton--proton collisions are sensitive to the production mechanisms of heavy quarks and to their fragmentation as well as hadronisation processes.
The measurement of the azimuthal-correlation function of prompt D mesons with charged particles in proton--proton collisions at a centre-of-mass energy of $\s = 13$ \TeV with the ALICE detector is reported, considering $\Dzero$, $\Dplus$, and $\Dstar$ mesons in the transverse-momentum interval $3 < \pt < 36$ \GeVc at midrapidity ($|y| < 0.5$), and charged particles with $\pt > 0.3$ \GeVc and pseudorapidity $|\eta| < 0.8$.
This measurement has an improved precision and provides an extended transverse-momentum coverage compared to previous ALICE measurements at lower energies.
The study is also performed as a function of the charged-particle multiplicity, showing no modifications of the correlation function with multiplicity within uncertainties.
The properties and the transverse-momentum evolution of the near- and away-side correlation peaks are studied and compared with predictions from various Monte Carlo event generators.
Among those considered, PYTHIA8 and POWHEG+PYTHIA8 provide the best description of the measured observables. The obtained results can provide guidance on tuning the generators.
\end{abstract}
\end{titlepage}

\setcounter{page}{2} 


\section{Introduction}
\label{sec:Introduction}
The study of angular correlations of heavy-flavour (charm and beauty) particles in ultra-relativistic hadronic collisions allows the investigation of fundamental properties of quantum chromodynamics (QCD) in the heavy-flavour domain~\cite{Mangano:1991jk, Andronic:2015wma}.
In particular, the angular-correlation function between prompt D mesons and charged particles in proton--proton (pp) collisions is sensitive to the mechanisms of charm-quark production, fragmentation, and hadronisation into charm hadrons. The term “prompt” refers to D mesons originating both from direct charm-quark fragmentation and from the strong decay of excited charm resonances, and excludes D mesons produced from beauty-hadron weak decays.
The typical structure of the two-dimensional correlation function between ``trigger'' D mesons and ``associated'' charged particles, expressed in terms of the pseudorapidity difference ($\Deta = \eta_{\rm ch}-\eta_{\rm D}$) and azimuthal angle difference ($\Dphi = \varphi_{\rm ch}-\varphi_{\rm D}$), features a near-side (NS) peak, centred at ($\Dphi,\Deta) = (0,0)$, and an away-side (AS) peak at $\Dphi = \pi$ that is elongated along $\Deta$~\cite{ALICE:2016clc}. Both peaks sit on top of an approximately flat continuum extending over the full ($\Dphi,\Deta$) range.
The height and width of the two correlation peaks are sensitive to the charm-quark production mechanisms.

Due to their large mass, the production of charm-quark pairs occurs through hard parton--parton scattering processes with large momentum transfers, and can be described by perturbative QCD (pQCD) calculations.
While in leading-order (LO) processes the two quarks are produced back-to-back in azimuth, the next-to-leading-order (NLO) production mechanisms, such as flavour excitation and gluon splitting, can break this topology and alter the shape of the two correlation peaks~\cite{Norrbin:2000zc}.
Recent studies at the LHC suggest a relevant contribution from gluon splitting to heavy-quark production, possibly underestimated by Monte Carlo (MC) event generators with LO or NLO accuracy~\cite{Aaij:2012dz,Khachatryan:2011wq}.
The analysis of the properties of the near-side peak also allows for detailing the fragmentation and hadronisation processes which, starting from a coloured charm quark, lead to the formation of a D meson surrounded by a spray of colourless particles, experimentally identifiable as a charm jet.
The production cross section of jets containing D mesons, as well as the jet momentum fraction carried by the D meson along the jet-axis direction, were recently measured by the ALICE and ATLAS Collaborations~\cite{Acharya:2019zup,Aad:2011td}.
In this regard, a systematic and differential analysis of the near-side correlation peak in terms of transverse momenta of trigger D meson ($\ptD$) and other associated fragmenting particles ($\ptass$), and of the angular distance of associated particles from the D mesons, can provide additional information with respect to measurements that treat charm jets as a whole entity.

In recent years, the study of high-multiplicity pp collisions has become of particular interest. The ALICE Collaboration has measured a faster-than-linear increase of prompt D-meson self-normalised yields for increasing relative event multiplicity in pp collisions at a centre-of-mass energy of $\s = 7$ TeV, employing both central- and forward-rapidity multiplicity estimators~\cite{Adam:2015ota}. A similar behaviour was also seen for open-beauty and hidden-charm hadrons, pointing towards sensitivity of heavy-quark production processes to event multiplicity.
Complementary information on a possible dependence of charm-quark fragmentation and hadronisation on event multiplicity can be provided by the study of D-meson and charged-particle azimuthal correlations as a function of the event multiplicity, searching in particular for modifications of the near-side peak structure.
This measurement is also crucial to validate the assumptions adopted to measure the elliptic-flow coefficient of D mesons and heavy-flavour decay muons in high-multiplicity pp collisions at $\s = 13$ TeV by CMS~\cite{CMS:2020qul} and ATLAS~\cite{ATLAS:2019xqc}, respectively. In these measurements, the elliptic-flow coefficient is extracted from two-particle correlation function of such heavy-flavour particles with charged particles, and the short-range correlation peaks related to heavy-quark fragmentation are removed from the correlation function exploiting low-multiplicity events, assuming independence of the correlation peaks from the event multiplicity.

MC event generators, like PYTHIA~\cite{Sjostrand:2006za,Sjostrand:2007gs}, EPOS~\cite{Werner:2010aa,Drescher:2000ha}, or HERWIG~\cite{Bahr:2008pv,Bellm:2015jjp,Bellm:2019zci}, or pQCD calculations coupled to parton-shower software, such as POWHEG~\cite{Nason:2004rx,Frixione:2007vw}, are widely used to reproduce ultra-relativistic hadronic collisions and provide predictions for a wide variety of physics observables.
As discussed in Ref.~\cite{Acharya:2019icl}, depending on the treatment of the various collision stages and implementation of specific features in each generator, such as hard parton--parton scattering matrix elements, parton-showering model, hadronisation algorithm, and underlying event generation, different predictions for D-meson and charged-particle correlation function will be obtained. A comparison of the predicted features of the correlation observables, in particular the peak yields and widths, with data measurements, can allow for validating and setting constraints to the MC generators~\cite{Prino:2016cni}.

A proper understanding of heavy-flavour correlations in pp collisions is also crucial in view of future studies in ultra-relativistic heavy-ion collisions.
In the first stages of such collisions a deconfined state of strongly-coupled matter, the quark--gluon plasma (QGP), is created. While traversing the medium, heavy quarks interact with the QGP constituents through radiative and collisional processes~\cite{Braaten:1991we,Baier:1996sk}, losing energy and having their original directions modified. This is expected to lead to a modification of the angular-correlation function between final-state heavy-flavour hadrons and associated charged particles, with respect to that observed in pp collisions.
Quantifying such modifications allows for investigation of specific properties of the QGP and its dynamics~\cite{Andronic:2015wma}.
In particular, the angular-correlation function is sensitive to the relative contributions of the two energy-loss processes, and can shed light on the path-length dependence of energy loss~\cite{Nahrgang:2013saa,Cao:2017hpp,Renk:2013xaa,Beraudo:2014boa}. Some first indications in this direction were provided by a recent measurement of $\Dzero$-meson and charged-hadron angular correlations in gold--gold collisions at a centre-of-mass energy per nucleon pair of $\snn = 200$ GeV performed by the STAR Collaboration at RHIC~\cite{Adam:2019tql}.
Validating MC event generators against the correlation function of heavy-flavour particles measured in pp collisions is thus fundamental for a correct understanding of the same observables that will be measured in the future in lead--lead (Pb--Pb) collisions at the LHC.

In this article, ALICE measurements of azimuthal correlations of prompt $\Dzero$, $\Dplus$, and $\Dstar$ mesons, together with their charge conjugates, with associated charged particles in pp collisions at $\s = 13$ TeV at midrapidity are reported. For prompt $\Dzero$-meson triggers the results of the correlation analysis are also reported as a function of the charged-particle event multiplicity, measured at forward and backward rapidity.
With respect to previous ALICE publications in pp collisions~\cite{ALICE:2016clc,Acharya:2019icl}, the new measurements extend the $\pt$ range of D mesons up to 36 \GeVc, and significantly improve the precision of the measured observables in the common $\pt$ range. Additionally, the measurements presented here, along with previous ALICE results at $\s = 5.02$ and 7 TeV, enable the study of the possible evolution of the correlation distributions and of the peak features as a function of the collision centre-of-mass energy.

The article is organised based on the following structure. Section~\ref{sec:ALICE_and_Data} describes the ALICE apparatus, as well as the data and MC samples used for this study. Section~\ref{sec:Analysis} highlights the procedure followed to obtain the azimuthal correlation function and to extract physical observables from it. In Section~\ref{sec:Systematics}, the sources of systematic uncertainties affecting the results are detailed. In Section~\ref{sec:Results}, the analysis results are presented and discussed, and a comparison with various model predictions is shown. Further model studies highlighting the specific contributions to the correlation function from initial- and final-state radiation and multi-parton interactions are reported in Section~\ref{sec:Detailed_sim}. A summary of the paper and its physics message is outlined in Section~\ref{sec:Summary}. 

\section{ALICE detector, data, and MC samples}
\label{sec:ALICE_and_Data}
A complete description of the ALICE detector and its performance can be found in Refs.~\cite{Aamodt:2008zz,Abelev:2014ffa}. The reconstruction of D mesons and charged particles was performed using detectors installed in the central barrel, with a pseudorapidity coverage of $|\eta| < 0.9$ and a magnetic field of $B = 0.5$ T parallel to the beam axis. In particular, the Inner Tracking System (ITS)~\cite{Aamodt:2010aa} and the Time Projection Chamber (TPC)~\cite{Alme:2010ke} were employed for the reconstruction of charged tracks and of primary and secondary vertices. The TPC, together with the Time-of-Flight (TOF) detector~\cite{Akindinov:2013tea}, also provided charged-particle identification (PID) information. The analysis also relied on detectors located along the beam line, at forward and backward rapidity. The V0 detector~\cite{Abbas:2013taa} is a set of scintillators covering the pseudorapidity ranges $2.8 < \eta < 5.1$ (V0A) and
$-3.7 < \eta < -1.7$ (V0C), used for triggering, background-event rejection, and event-multiplicity estimation. The T0 detector is an array of Cherenkov counters, located along the beam line, at a distance of $+370$ cm (T0A) and $-70$ cm (T0C) from the nominal interaction point, and provides the collision starting time needed by the TOF~\cite{Adam:2016ilk}.

The data sample used for the analysis consisted of pp collisions at $\s = 13$ TeV collected during the 2016, 2017, and 2018 data taking periods, with a total integrated luminosity of about 29 nb$^{-1}$, based on the visible cross section measured with the V0 detector~\cite{ALICE-PUBLIC-2016-002}. 
The collisions were recorded if they satisfied a minimum-bias trigger, requiring the presence of signals in both V0 detectors in coincidence with a bunch crossing in the ALICE interaction region. This trigger was fully efficient for selecting events containing D mesons with $\pt > 1$ \GeVc.
Contamination of tracks from pile-up events (multiple collisions occurring in the same bunch crossing) was suppressed by discarding events where multiple primary vertices were reconstructed with the Silicon Pixel Detector (SPD), which constitutes the first two layers of the ITS.
Timing information provided by the V0, as well as the correlation between the number of hits and the number of track segments in the SPD, were employed to reject beam--gas interactions.
Only events with a reconstructed primary vertex within $\pm 10$ cm from the nominal centre of the ALICE detector along the beam direction were considered to grant a uniform acceptance for the central-barrel detectors.

The multiplicity-differential analysis was performed in four independent multiplicity classes, defined in terms of the total energy deposit in the V0 detectors by charged particles passing through them (V0M amplitude).
The rapidity gap between the V0 detectors and the central barrel in which the D mesons and charged particles were reconstructed assured the absence of significant auto-correlations between the correlation peaks and the multiplicity estimate.
The V0M amplitudes were converted to percentiles of the inelastic collisions with at least one charged particle produced in $|\eta| < 1$ (INEL$>$0), corresponding to about 75\% of the total inelastic cross section, as described in Ref.~\cite{Acharya:2020kyh}.

The corresponding INEL$>$0 percentiles ($\sigma$/$\sigma_{\rm INEL>0}$) of the four V0M multiplicity classes are reported in Table~\ref{tab:V0Mclass}, together with the related average number of charged particles, $\langle {\rm d}N_{\rm ch}/{\rm d}\eta \rangle$ in $|\eta| < 0.5$.
A specific high-multiplicity trigger (V0HM) was used for the V0M multiplicity class I, to enhance the statistical precision of this particular class. The V0HM trigger recorded only events with a multiplicity large enough to pass a threshold of V0M amplitude. This trigger covered the whole span of V0M multiplicity class I. Only the data periods granting a uniform efficiency of the V0HM trigger inside the range covered by V0M multiplicity class I were considered for the multiplicity-differential analysis, resulting in an integrated luminosity specific to the V0HM trigger of about 7.7 pb$^{-1}$.

\begin{table}[h]
\centering
\caption{The percentiles of the INEL$>$0 cross section of the four V0M-based event-multiplicity classes and the corresponding midrapidity charged-particle multiplicities. Systematic uncertainties on the charged-particle multiplicity values, derived from~\cite{Acharya:2020kyh}, are also reported.}
\begin{tabular}{c|c|c|c|c}
\hline
\textbf{V0M multiplicity class} & \textbf{I} & \textbf{II} & \textbf{III} & \textbf{IV} \\
\hline
$\sigma$/$\sigma_{\rm INEL>0}$ (\%) & 0--0.0915 & 0.0915--9.149 & 9.149--27.50 & 27.50--100 \\
$\langle {\rm d}N_{\rm ch}/{\rm d}\eta \rangle$ & 31.15 $\pm$ 0.40 & 18.39 $\pm$ 0.23 & 11.46 $\pm$ 0.15 & 4.41 $\pm$ 0.06 \\
\hline
\end{tabular}
\label{tab:V0Mclass}
\end{table}

To evaluate the corrections to the azimuthal-correlation measurements, several MC simulations of pp collisions at $\s = 13$ TeV were used, produced with the PYTHIA 6.4.25 event generator~\cite{Sjostrand:2006za} with the Perugia-2011 tune~\cite{Skands:2010ak}.
For the corrections specific to D mesons, a sample of pp collisions was produced with the same generator, with each event containing either a ${\rm c}\overline{\rm c}$ or a ${\rm b}\overline{\rm b}$ pair in the rapidity range $[-1.5,1.5]$.
In addition, simulated events satisfying a minimum threshold of midrapidity charged-particle multiplicity were employed, as they provided sufficient statistical precision to evaluate the corrections for the V0M multiplicity class I.
The simulations included the full description of the detector geometry, response, and conditions during the data taking via the GEANT3 package~\cite{Brun:1082634}.

\section{Analysis overview}
\label{sec:Analysis}
The procedure for the evaluation of the D-meson (intending, in the context of this study, $\Dzero$, $\Dstarwide$, and $\Dplus$ mesons) and charged-particle azimuthal correlation function and the related corrections is described in Section~\ref{sec:Analysis_Int} for the multiplicity-integrated analysis. The multiplicity-differential analysis largely followed the same approach, although some of the quantities and the corrections were evaluated independently in each V0M multiplicity class, or with a slightly modified procedure. Such differences are highlighted in Section~\ref{sec:Analysis_Diff}.
The extraction of physical observables from the fully-corrected correlation function, in common between the two studies, is discussed in Section~\ref{sec:Analysis_Fit}.

\subsection{Evaluation and correction of the azimuthal correlation function for the multiplicity-integrated analysis}
\label{sec:Analysis_Int}
All stages of the analysis were mostly unaltered with respect to those performed for the same study in pp collisions at $\s = 5.02$ TeV, and are comprehensively described in Ref.~\cite{Acharya:2019icl}. Thus, they will be only briefly summarised in the following.

$\Dzero$-, $\Dstarwide$-, and $\Dplus$-meson candidates and their charge conjugates, used as trigger particles in the analysis, were reconstructed from their hadronic decay channels ${\Dzero \to {\rm K}^{-}\pi^{+}}$, with branching ratio BR~=~(3.95 $\pm$ 0.03)\%, ${\Dplus \to {\rm K}^{-}\pi^{+}\pi^{+}}$ with BR~=~(9.38 $\pm$ 0.16)\%, and ${\Dstar \to \Dzero\pi^{+}} \to {\rm K}^{-}\pi^{+}\pi^{+}$ with BR~=~(2.67 $\pm$ 0.03)\%~\cite{Zyla:2020zbs}. A topological selection, exploiting the characteristic displacement of the D-meson decay vertices with respect to the primary vertex, and particle-identification information on the D-meson decay products were employed to suppress the combinatorial background. Further details on the selection are provided in Ref.~\cite{Acharya:2019mgn}. The same criteria were followed in this analysis, apart from an optimisation of the selection values performed on the specific samples used to further increase the signal-to-background ratio of D-meson candidates. A fit to the invariant-mass distribution of selected D-meson candidates was performed as described in Ref.~\cite{Acharya:2019mgn}, in order to extract the D-meson yield, $S_{\mathrm{peak\ region}}$, in a $\pm2 \sigma$ region from the centre of the invariant-mass signal peak, where $\sigma$ is the width of the Gaussian component of the fit function describing the signal peak.
The associated particles included charged primary~\cite{ALICE-PUBLIC-2017-005} pions, kaons, protons, electrons, and muons with $\ptass > 0.3$ \GeVc and $|\eta| < 0.8$. The decay products of the trigger D meson were excluded from the associated-particle sample.
The tracks reconstructed in the ALICE central barrel were accepted as associated particles if they satisfied selection criteria based on the quality of their reconstruction in the ITS and TPC detectors, as detailed in  Ref.~\cite{ALICE:2016clc}. Additionally, a maximum distance of closest approach (DCA) of the track to the primary vertex of 1 cm in the transverse ($xy$) plane and along the beam line ($z$-direction) was required. This selection suppressed the contamination of non-primary particles to about 5\% for $0.3 < \ptass < 1$ \GeVc, falling to below 1\% for $\ptass > 2$ \GeVc.
As a result of the applied selection criteria, the associated-track reconstruction and selection efficiency ranged from about $80\%$ for $\ptass = 0.3$ \GeVc up to about $90\%$, increasing with $\ptass$.

Selected D-meson candidates within $\pm2 \sigma$ from the centre of the invariant-mass signal peak (``peak region'') were correlated with associated particles reconstructed and selected in the same event. A two-dimensional angular-correlation function, $C(\Dphi, \Delta\eta)_{\mathrm{peak\ region}}$, was built for each of the five D-meson $\pt$ intervals, ranging from 3 to 36 \GeVc, and for the associated track $\pt$ interval $\ptass > 0.3$~\GeVc and its sub-ranges: $0.3 < \ptass < 1$~\GeVc, $1 <\ptass< 2$~\GeVc, and $2 <\ptass < 3$~\GeVc.
The limited detector acceptance and efficiency for the reconstruction and selection of D-meson candidates and associated particles were accounted for by weighting each correlation pair by 1/$(A\times\epsilon)^{\rm assoc} \times 1$/($A\times\epsilon)^{\rm trig}$, where $A$ and $\epsilon$ represent the acceptance and efficiency factors, respectively, evaluated using MC simulations. 
The $(A \times\epsilon)^{\rm trig}$ values were dependent on the event multiplicity. In particular, the D-meson selection efficiency decreased for low-multiplicity events, due to the degraded resolution on the primary-vertex position, which enters into several topological selection criteria. To account for this dependency, the $(A \times\epsilon)^{\rm trig}$ weights were evaluated and applied in narrow intervals of SPD tracklet multiplicity.
The entries of the invariant-mass distributions of D-meson candidates were also scaled by 1/($A\times\epsilon)^{\rm trig}$ to allow a correct per-trigger normalisation of the correlation function, as detailed later.
Additional losses due to pair acceptance effects were taken into account by applying a mixed-event correction. Specifically, D-meson candidate triggers were correlated with associated charged particles from other events with similar midrapidity event multiplicity and primary vertex location along the beam axis. Track segments reconstructed by associating hits in the two SPD layers and pointing to the reconstructed primary vertex (SPD tracklets) were used as the midrapidity multiplicity estimator for the event classification. In this way, a mixed-event correlation function, $\mathrm{ME}(\Dphi, \Delta\eta)_{\mathrm{peak\ region}}$, was built and used to weight the same-event correlation function $C(\Dphi, \Delta\eta)_{\mathrm{peak\ region}}$.

The correlation function $C(\Dphi, \Delta\eta)_{\mathrm{peak\ region}}$ also included a contribution from background D-meson candidates. This contribution was statistically removed by employing a sideband-subtraction technique. A sideband-region correlation distribution, $C(\Dphi,\Delta\eta)_{\rm{sidebands}}$, was built by considering as trigger particles $\Dzero$- and $\Dplus$-meson candidates 4--8$\sigma$ from the centre of the invariant-mass signal peak, in both directions, and $\Dstar$-meson candidates 5--10$\sigma$ to the right of the invariant-mass signal peak centre. A mixed-event correction was applied to the sideband-region correlation function, following the same procedure described above for $C(\Dphi, \Delta\eta)_{\mathrm{peak\ region}}$. Subsequently, the sideband-region correlation function was subtracted from that of the peak region to obtain the signal correlation function, $C(\Dphi, \Delta\eta)_{\mathrm{signal}}$.
The above procedure is described in Eq.~\ref{eq:Proc1}
\begin{equation}
\label{eq:Proc1}
\centering
    \Tilde{C}_{\rm{signal}}(\Dphi, \Delta\eta) = \frac{1}{S_\mathrm{peak\ region}}\left( \frac{C(\Dphi, \Delta\eta)}{\mathrm{ME}(\Dphi, \Delta\eta)} \bigg|_\mathrm{peak\ region} - \frac{B_\mathrm{peak\ region}}{B_\mathrm{sidebands}}\frac{C(\Dphi, \Delta\eta)}{\mathrm{ME}(\Dphi, \Delta\eta)} \bigg|_\mathrm{sidebands}\right), 
\end{equation}

where the factor $1/S_{\mathrm{peak\ region}}$ provides a per-trigger normalisation to the signal correlation function, as denoted by a $\Tilde{C}$ symbol. The terms $B_{\mathrm{peak\ region}}$ and $B_{\rm sidebands}$ quantify the number of background D-meson candidate triggers in the invariant-mass peak region and sidebands, respectively.

The two-dimensional correlation function $\Tilde{C}_{\rm{signal}}(\Dphi,\Delta\eta)$ was integrated in the range $|\Delta\eta| < 1$, obtaining the per-trigger azimuthal correlation function $\Tilde{C}_{\rm{signal}}(\Dphi)$, in order to grant sufficient statistical precision. For the same reason, due to its symmetry around $\Dphi = 0$ and $\Dphi = \pi$, the correlation function was restricted to the $0 \leq \Dphi \leq \pi$ interval, averaging the symmetric points in the ranges [0, $\pi$] and [$-\pi$, 0].

The residual contamination of non-primary associated tracks satisfying the DCA selection criteria was removed by multiplying  $\Tilde{C}_{\rm{signal}}(\Dphi)$ by a $\Dphi$-differential purity correction factor $p_{\rm prim}(\Dphi)$, evaluated from MC simulations. A slight increase with $\ptass$ was observed for the $\Dphi$-averaged value of this factor, which ranged between 0.94 and 0.99. Modulations as a function of $\Dphi$ up to 2\% were obtained.
The fraction of prompt D mesons ($f_{\rm prompt}$) accounted for approximately $90\%$ of D mesons accepted by the topological and PID selection, with a slight increase for increasing $\pt$. The remaining contribution was composed of D mesons produced by beauty-hadron decays (feed-down D mesons). Thus, the azimuthal correlation function $\Tilde{C}_{\rm{signal}}(\Dphi)$ included a contribution from feed-down D-meson candidate triggers.
For small $\Dphi$ values, the shape of this contribution was distorted by the topological selection applied to the D-meson candidates, which was more efficient in selecting beauty-hadron decays featuring a small opening angle between the D-meson candidate trigger and the other decay particles.
The natural shape of the feed-down contribution to the azimuthal correlation function was recovered by evaluating the amount of the distortion via MC studies, and applying a correction factor $b_{\rm B\mhyphen bias}(\Dphi)$ to the data correlation function, as explained in detail in Ref.~\cite{Acharya:2019icl}. The correction amounted to a maximum of 4.5\% for $\Dphi = 0$ and was substantially smaller for larger $\Dphi$ values.
After applying this correction, the feed-down contribution to the measured correlation function was removed as follows. A template of the per-trigger azimuthal correlation function from feed-down D-meson triggers, $\Tilde{C}^{\rm MC\ templ}_{\rm{feed\mhyphen down}}(\Dphi)$, was evaluated for each $\pt$ range with the PYTHIA6 event generator with Perugia-2011 tune at generator level (i.e.~without detector effects and selection criteria). After being rescaled to the expected fraction of feed-down D-meson candidate triggers, $1-f_{\rm prompt}$, this contribution was subtracted from the purity-corrected azimuthal-correlation function. The fully-corrected, per-trigger azimuthal correlation function of prompt D mesons and charged particles was obtained with this procedure, as summarised in Eq.~\ref{eq:Proc2}

\begin{equation}
\label{eq:Proc2}
\centering
    \frac{1}{N_{\rm D}} \times \frac{{\rm d}N^{\rm assoc}}{{\rm d}\Dphi}(\Dphi) =
    \frac{1}{f_{\rm prompt}} \left[b_{\rm B\mhyphen bias}(\Dphi)\times p_{\rm prim}(\Dphi)\times \Tilde{C}_{\rm{signal}}(\Dphi, \Delta\eta) - \left(1-f_{\rm prompt}\right)\Tilde{C}^{\rm MC\ templ}_{\rm{feed\mhyphen down}}(\Dphi) \right].
\end{equation}

\subsection{Details specific to the multiplicity-differential analysis}
\label{sec:Analysis_Diff}
For the multiplicity-differential analysis, only $\Dzero$ mesons and their charge conjugates were used as trigger particles. The same selection criteria chosen for the multiplicity-integrated case were used for the $\Dzero$-meson candidate triggers and the associated charged particles.
The evaluation of the two-dimensional correlation function for the peak region and the sidebands of the $\Dzero$-meson invariant-mass distributions was performed independently in each of the four V0M multiplicity classes, considering the same $\pt$ ranges of the multiplicity-integrated analysis, with the exception of $24 < \ptD < 36$ \GeVc, where the amount of collected data was not sufficient for performing the study.
The invariant-mass distributions of the V0M multiplicity class I were characterised by a larger data sample and a larger statistical significance of the $\Dzero$ mass peak, profiting from the usage of the V0HM trigger, although they also showed a lower signal-to-background ratio due to the enhanced underlying-event activity.

For each V0M multiplicity class, the per-trigger azimuthal correlation function was obtained following Eq.~\ref{eq:Proc1} and Eq.~\ref{eq:Proc2}, but some of the quantities entering these equations were evaluated with a modified procedure.
For the mixed-event correction, a different classification of the events in terms of SPD tracklet multiplicity needed to correlate $\Dzero$ mesons and charged particles from events with similar features was considered for each V0M multiplicity class, since the SPD tracklet multiplicity distributions obtained for each V0M multiplicity class were significantly different.
The MC events used to evaluate the values of 1/($A\times\epsilon)^{\rm trig}$, for each V0M multiplicity class, were reweighted in order to reproduce the corresponding SPD tracklet multiplicity measured in data.
The same values of charged-particle reconstruction and selection efficiency were instead used for the four V0M multiplicity classes since a negligible dependence of the efficiency on the event multiplicity was found in previous ALICE studies in the same collision system~\cite{Acharya:2020zji}.

The purity correction, $p_{\rm prim}(\Dphi)$, was evaluated independently for each V0M multiplicity class, by applying the same MC reweighting procedure used for the $\Dzero$-meson efficiency evaluation. A very small dependence on event multiplicity was obtained, with the overall differences smaller than $0.5 \%$ between the values obtained in the four V0M multiplicity classes and those evaluated for the multiplicity-integrated analysis.
Similarly, the correction factor $b_{\rm B\mhyphen bias}(\Dphi)$ was estimated separately for each of the four V0M multiplicity classes. A slight increase of the correction of about 1--2\% depending on the $\pt$ range was obtained with decreasing event multiplicity. The largest value of $b_{\rm B\mhyphen bias}(\Dphi)$ was about $5.5\%$ for V0M multiplicity class IV at $\Dphi=0$ for the lowest $\ptD$ interval and largest $\ptass$ interval.
The feed-down subtraction procedure was left unaltered for the evaluation of the central values of the correlation function, assuming no modification of the prompt $\Dzero$-meson fraction and beauty-quark fragmentation with event multiplicity. However, as described in more detail in Section~\ref{sec:Systematics}, an additional systematic uncertainty was considered, accounting for a possible variation of the feed-down-to-prompt $\Dzero$-meson production ratio with event multiplicity, which would impact the value of $f_{\rm prompt}$.

\subsection{Quantitative evaluation of correlation peak features}
\label{sec:Analysis_Fit}
Based on their consistency within uncertainties, an average of the azimuthal-correlation functions for $\Dzero$, $\Dplus$, and $\Dstarwide$ meson triggers was evaluated for the multiplicity-integrated analysis. The average was obtained by weighting the correlation function from each species by the inverse of the quadratic sum of its statistical and systematic uncertainties uncorrelated among the three D-meson species.
For each $\pt$ interval, the averaged azimuthal correlation function was fitted with the following function

\begin{equation}
\label{eq:fit}
    f(\Dphi) = a + \frac{Y_{\rm NS}\times\beta}{2\alpha \Gamma(1/\beta)}\times e^{-\left(\frac{\Dphi}{\alpha}\right)^\beta} + \frac{Y_{\rm AS}}{\sqrt{2\pi}\sigma_{\rm AS}}\times e^{-\frac{(\Dphi - \pi)^2}{2\sigma^2_{\rm AS}}}.
\end{equation}

This function is composed of a generalised-Gaussian component for the description of the near-side peak (with the mean fixed at $\Dphi = 0$), a Gaussian component for the away-side peak (with the mean fixed at $\Dphi = \pi$), and a constant term (baseline) to account for the flat contribution that lies beneath the two correlation peaks. To grant sufficient stability to the fit, the $\beta$ parameter of the generalised Gaussian was fixed to the value obtained for the correlation distribution predicted by PYTHIA8 simulations at generator level.
In Eq.~\ref{eq:fit}, the baseline value $a$ was fixed to the weighted average of the points in the range $\pi/4 < |\Dphi| < \pi/2$ (transverse region), to reduce the fit sensitivity to statistical fluctuations. The inverse of the squared statistical uncertainties of the points were used as weights.
The fit to the correlation function allowed the extraction of quantitative observables that characterise the correlation peaks. In particular, the near- and away-side peak yields were obtained as the integral of the components describing each correlation peak, and their widths were parameterised by the quantities $\alpha\sqrt{\Gamma(3/\beta)/\Gamma(1/\beta)}$ (square root of the generalised-Gaussian variance) and $\sigma_{\rm AS}$, respectively.

\section{Systematic uncertainties}
\label{sec:Systematics}

\subsection{Systematic uncertainties for the multiplicity-integrated analysis}
\label{sec:Systematics_MultInt}
The azimuthal correlation function obtained from the multiplicity-integrated analysis is affected by several systematic uncertainties due to the specific procedure and assumptions introduced for its evaluation.  In the following, the approach used to estimate each systematic uncertainty source is briefly described.

The evaluation of $S_{\mathrm{peak\ region}}$ and $B_{\mathrm{peak\ region}}$ from the fit to the D-meson invariant-mass distributions (as described in Section~\ref{sec:Analysis_Int} and Eq.~\ref{eq:Proc1}) introduced a systematic uncertainty on the correlation function. The uncertainty was estimated by varying the fit procedure, specifically, by modelling the background distribution with a linear function or a second-order polynomial function instead of an exponential function (for $\Dzero$ and $\Dplus$ mesons only, where there is not a straightforward choice for the background fit function), considering a different histogram binning, varying the fit range, fixing the mean of the Gaussian term describing the mass peak to the world-average D-meson mass~\cite{Zyla:2020zbs}, or fixing the Gaussian width to the value obtained from MC studies. A systematic uncertainty ranging from 0.5\% to 1.5\%, depending on the $\ptD$ range and similar for all D-meson species, was estimated from the corresponding variation of the azimuthal-correlation function. No dependence on $\Delta\varphi$ was found.

A 0.5\% to 2\% systematic uncertainty, depending on $\ptD$ range and D-meson species, was introduced due to the possible dependence of the shape of the background correlation function on the invariant-mass value of the trigger D meson. This source of uncertainty was estimated by evaluating $\Tilde{C}_{\rm{sidebands}}(\Dphi, \Delta\eta)$ considering different invariant-mass sideband ranges. For $\Dzero$ and $\Dplus$ mesons, for which a sideband was defined on each side of the invariant-mass signal peak, $\Tilde{C}_{\rm{sidebands}}(\Dphi, \Delta\eta)$ was also evaluated considering only the left or the right sideband. No azimuthal dependence was observed for this uncertainty.

The evaluation of the associated-particle reconstruction efficiency via MC studies introduced a further systematic uncertainty, estimated by varying the quality selection criteria applied on the reconstructed tracks, i.e.~removing or tightening the request on minimum number of ITS clusters, requiring a hit on at least
one of the two SPD layers, or varying the request on the minimum number of space points reconstructed in the TPC. An uncertainty of 4\% to 5\% was estimated, independent of the D-meson species, and no significant trend in $\Delta\varphi$ was observed.

A systematic uncertainty affecting the D-meson reconstruction efficiency, related to potentially different distributions of the topological variables in MC and data, was estimated by testing a set of tighter and looser topological selections on D-meson candidates. An uncertainty ranging from 0.5\% to 2\% was assigned and the effect on the azimuthal correlation function was found to be $\Delta\varphi$ independent.

The correlation function has an uncertainty related to the evaluation of the residual contamination from secondary particles~\cite{ALICE-PUBLIC-2017-005}. To determine this, the analysis was repeated by varying the DCA selection in the $xy$ plane from 0.1 cm to 2.4 cm and re-evaluating the purity correction of primary tracks for each variation. This resulted in a maximum, $\Delta\varphi$-independent, systematic uncertainty of 2\% on the azimuthal-correlation function.

In addition to the above contributions, which all act as a scale uncertainty, the azimuthal correlation function is also affected by $\Delta\varphi$-dependent systematic uncertainties. The uncertainty on the evaluation of the beauty feed-down contribution to the azimuthal correlation function was determined by employing alternate templates of feed-down azimuthal-correlation functions, obtained from different event generators (PYTHIA6 with the Perugia-2010 tune~\cite{Skands:2010ak} and PYTHIA8 with the 4C tune~\cite{Corke:2010yf}), and by varying the value of $f_{\rm prompt}$ following the procedure described in Ref.~\cite{Acharya:2019mgn}. A $\Delta\varphi$-dependent uncertainty was obtained with a maximum value of 5\%. The near-side region for the feed-down D-meson component of the correlation function was affected by a bias, favouring topologies with a small opening angle between the D meson and the other beauty-hadron decay products. This was corrected for as discussed in Section~\ref{sec:Analysis_Int}, and a $\Delta\varphi$-dependent bilateral and symmetric uncertainty for a possible over- or under-correction of this bias was evaluated as detailed in Ref.~\cite{Acharya:2019icl}. The largest value of the uncertainty was found to be 2.5\% for $\Delta\varphi \approx 0$.

The estimated systematic uncertainty values from each of the above sources affecting the azimuthal correlation function are summarised in Table~\ref{systematic_table}. The overall systematic uncertainty in each $\Delta\varphi$ bin of the correlation function was obtained as the sum in quadrature of the aforementioned contributions.

The systematic uncertainties on the peak observables were evaluated by considering several contributions:
(i) the impact on the physical observables induced by the baseline position was estimated by considering alternate $\Delta\varphi$ ranges for determining its value and repeating the fit; similarly, the possible bias induced by fixing the $\beta$ parameter of the near-side to the predicted PYTHIA8 value was estimated by allowing $\beta$ to vary within $\pm20\%$ of that value. For each observable, the root-mean-square of the relative variations from these alternative fits with respect to the central value of the observable was considered; (ii) the impact of the $\Delta\varphi$-dependent uncertainty on the correlation function was accounted for by coherently shifting its points to the upper and lower edges of their $\Delta\varphi$-dependent systematic uncertainty values. The fit was repeated and the variation of each observable value with respect to the default value was considered in each direction; (iii) the overall $\Delta\varphi$-independent systematic uncertainty acts as a scaling factor on the correlation function, hence it impacted the near- and away-side peak yield values by the same relative amount.
The overall systematic uncertainty on the peak yields was therefore obtained by summing in quadrature the contributions from (i), (ii), and (iii). For the near-side and away-side widths, which are insensitive to scale factors, the sum in quadrature of only the contributions from (i) and (ii) was considered.

\subsection{Systematic uncertainties for the multiplicity-differential analysis}

Some of the systematic uncertainties affecting the correlation function were estimated separately in each of the V0M multiplicity class, following the same prescriptions described in Section~\ref{sec:Systematics_MultInt}.
In particular, this was done for the uncertainties on the yield extraction and background correlation shape since they are related to the features of the invariant-mass distributions, which show a significant multiplicity evolution.
The uncertainty related to the bias affecting the topological selection of feed-down $\Dzero$ mesons was also re-evaluated, since a slight multiplicity dependence of the related correction was found. Similar values of the uncertainty were obtained compared to the multiplicity-integrated case.
For the subtraction of the beauty feed-down contribution, an additional systematic uncertainty was considered, related to a possible multiplicity dependence of the relative fraction of feed-down $\Dzero$ mesons in the $\Dzero$-meson raw yields that determines the amount of the feed-down contribution. The evaluation of this uncertainty followed a similar procedure as the one described in Ref.~\cite{Adam:2015ota}, and led to an asymmetry of the feed-down systematic uncertainty, which increased up to $^{+5\%}_{-9\%}$ for the V0M multiplicity class I.
For the other systematic uncertainty sources affecting the azimuthal-correlation function, the same values estimated for the multiplicity-integrated analysis were adopted.
The uncertainty values for the multiplicity-differential analysis are reported in Table~\ref{systematic_table}.

\begin{table}[th]
\centering
 \caption{The list of the systematic uncertainty contributions affecting the azimuthal correlation function and their typical values. If not specified, the uncertainties do not depend on $\Delta\varphi$.}
 \begin{tabular}{l|c|c}
 \hline
 \textbf{Analysis}  & \textbf{Multiplicity-integrated} & \textbf{Multiplicity-differential} \\
 \hline
 Yield extraction & 0.5--1.5\% & 1--2\% \\
 Background $\Delta\varphi$ function & 0.5--2\% & 1--3\% \\
 Associated-track reconstruction efficiency & 4--5\% & 4--5\% \\
 D-meson efficiency & 0.5--2\% & 0.5--2\% \\
 Primary-particle purity & 1--2\% & 1--2\% \\
 Feed-down subtraction & $\leq$5\%, $\Delta\varphi$-dependent & $\leq^{+5\%}_{-9\%}$, $\Delta\varphi$-dependent \\
 Selection bias to feed-down contribution & $\leq$2.5\%, $\Delta\varphi$-dependent & $\leq$3\%, $\Delta\varphi$-dependent \\
 \hline
 \end{tabular}
 \label{systematic_table}
\end{table}

The evaluation of the systematic uncertainties on the near- and away-side peak observables was unmodified with respect to the procedure described in Section~\ref{sec:Systematics_MultInt} for the multiplicity-integrated analysis.
In addition, the impact on the peak observables related to the possible presence of long-range azimuthal correlations between $\Dzero$ mesons and charged particles in high-multiplicity collisions was studied by replacing in Eq.~\ref{eq:fit} the constant term with a $v_{2\Delta}$-like modulation. The elliptic-flow coefficient values of $\Dzero$ mesons and charged particles adopted to evaluate the modulation were defined based on the measurements in Ref.~\cite{CMS:2020qul} and Ref.~\cite{ALICE:2019zfl}, respectively. 
For the V0M multiplicity class I, variations within 2\% were found for all observables and $\pt$ ranges except for the range 3 $< \ptD <$ 5~\GeVc, where a reduction of the near- and away-side peak yields as large as 8\% was observed. These differences were assigned as a systematic uncertainty.

\section{Results}
\label{sec:Results}
\subsection{Multiplicity-integrated results in pp collisions at $\s = 13$~TeV}
\label{sec:Result_pp_vs_energies}
For the multiplicity-integrated analysis, the azimuthal correlation function of D mesons with charged particles was computed for the five D-meson $\pt$ intervals $3 < \ptD < 5$~\GeVc, $5 < \ptD < 8$~\GeVc, $8 <\ptD< 16$~\GeVc, $16 < \ptD < 24$~\GeVc, and $24 < \ptD < 36$~\GeVc, and for associated particle $\pt$ range $\ptass > 0.3$~\GeVc and the sub-intervals $0.3 < \ptass < 1$~\GeVc, $1 <\ptass< 2$~\GeVc, and $2 <\ptass < 3$~\GeVc.

\begin{figure}[t]
	\centering
	{\includegraphics[width=0.99\textwidth]{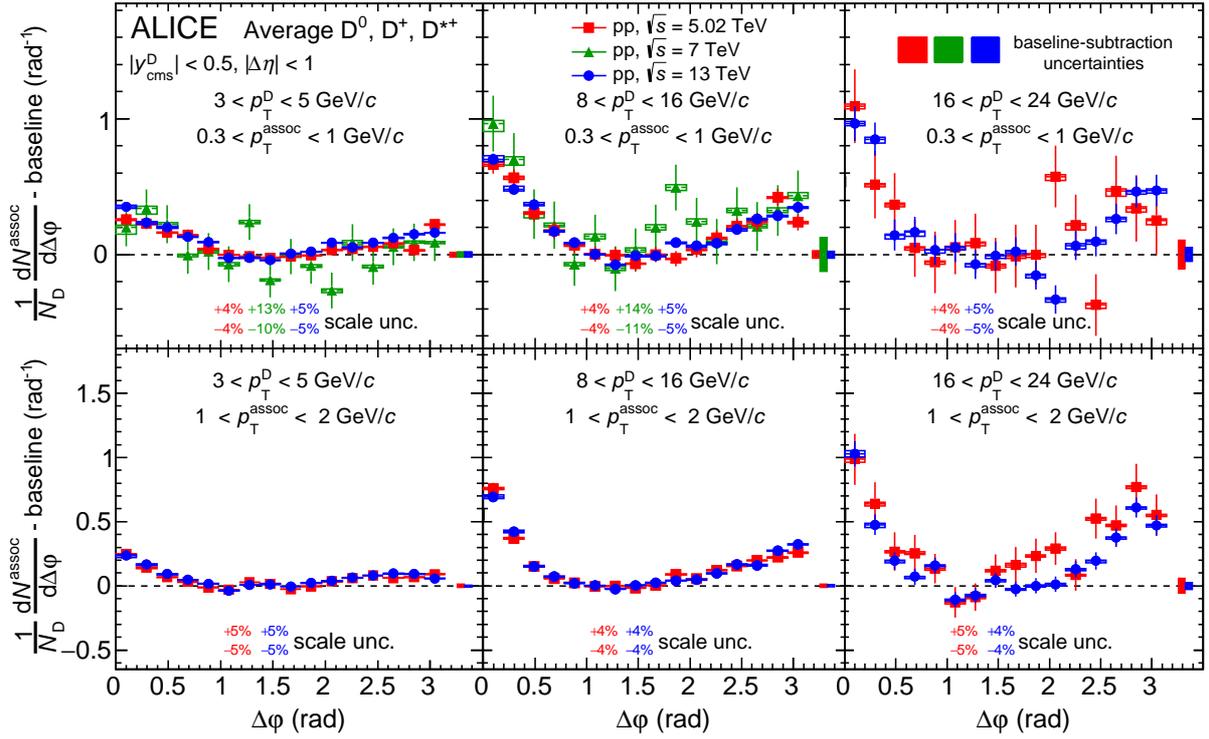} }%
	\caption {Average of the azimuthal-correlation functions of $\Dzero$, $\Dplus$, and $\Dstar$ mesons with associated particles, after the baseline subtraction, in pp collisions at $\sqrt{s}$ = 5.02~\cite{Acharya:2019icl}, 7~\cite{ALICE:2016clc}, and 13~TeV, for 3 $< \ptD <$ 5~\GeVc, 8~$< \ptD <$~16~\GeVc, and 16 $<\ptD<$ 24~\GeVc (from left to right) and 0.3 $<\ptass<$ 1~\GeVc, 1 $<\ptass<$ 2~\GeVc (top and bottom panels, respectively). Data at $\sqrt{s}$ = 7~TeV are not available for all the $p_{\mathrm{T}}$ regions. Statistical and $\Delta \varphi$-dependent systematic uncertainties are shown as vertical error bars and boxes, respectively, and $\Delta \varphi$-independent uncertainties are written as text. The uncertainties from the subtraction of the baseline are displayed as boxes at $\Delta \varphi>\pi$.}%
	\label{fig:Comparisons_Energy_DPhi}%
\end{figure}

\begin{figure}[t]
	\centering
	\includegraphics[width=0.46\textwidth]{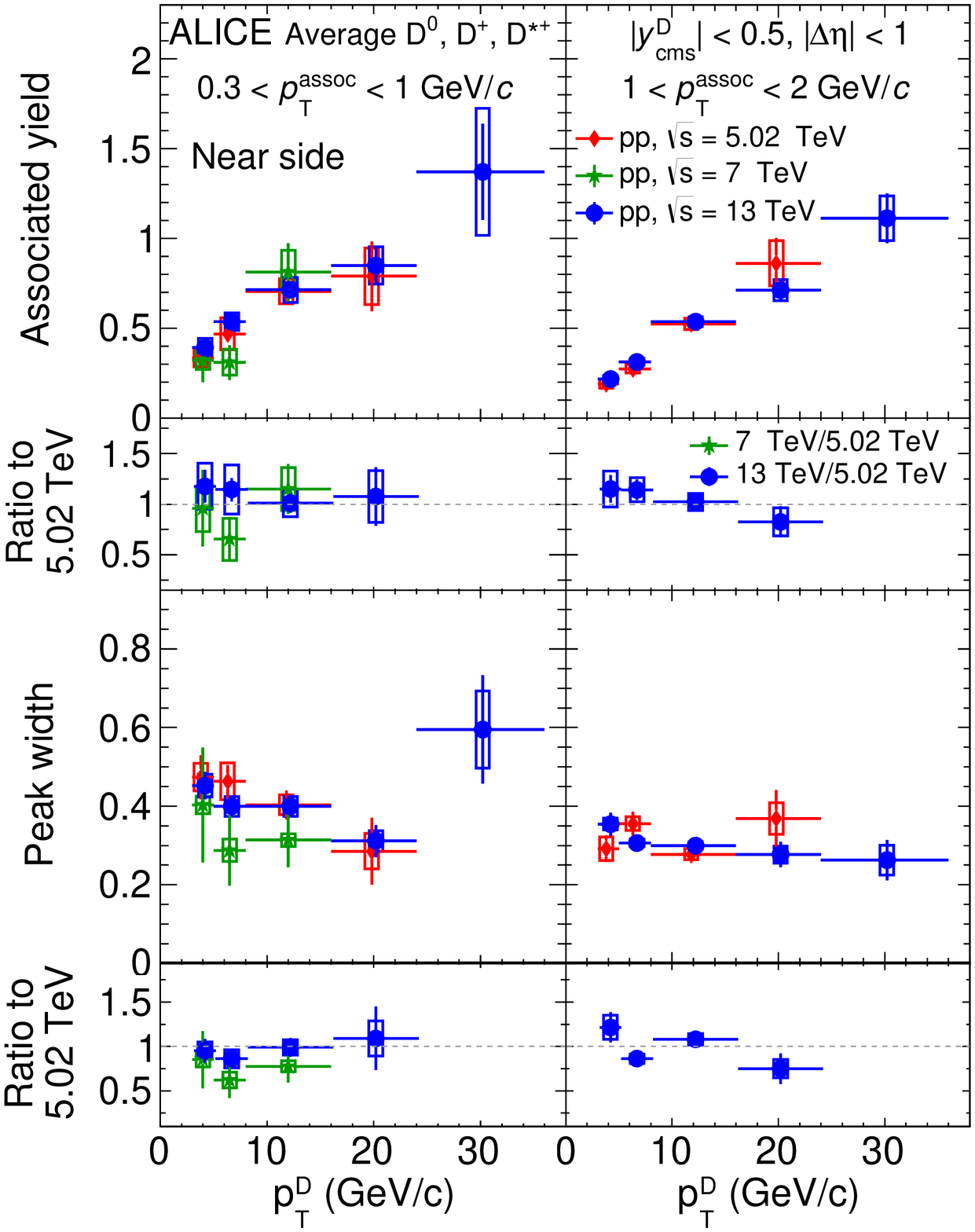}\hspace{0.2cm}
	\includegraphics[width=0.46\textwidth]{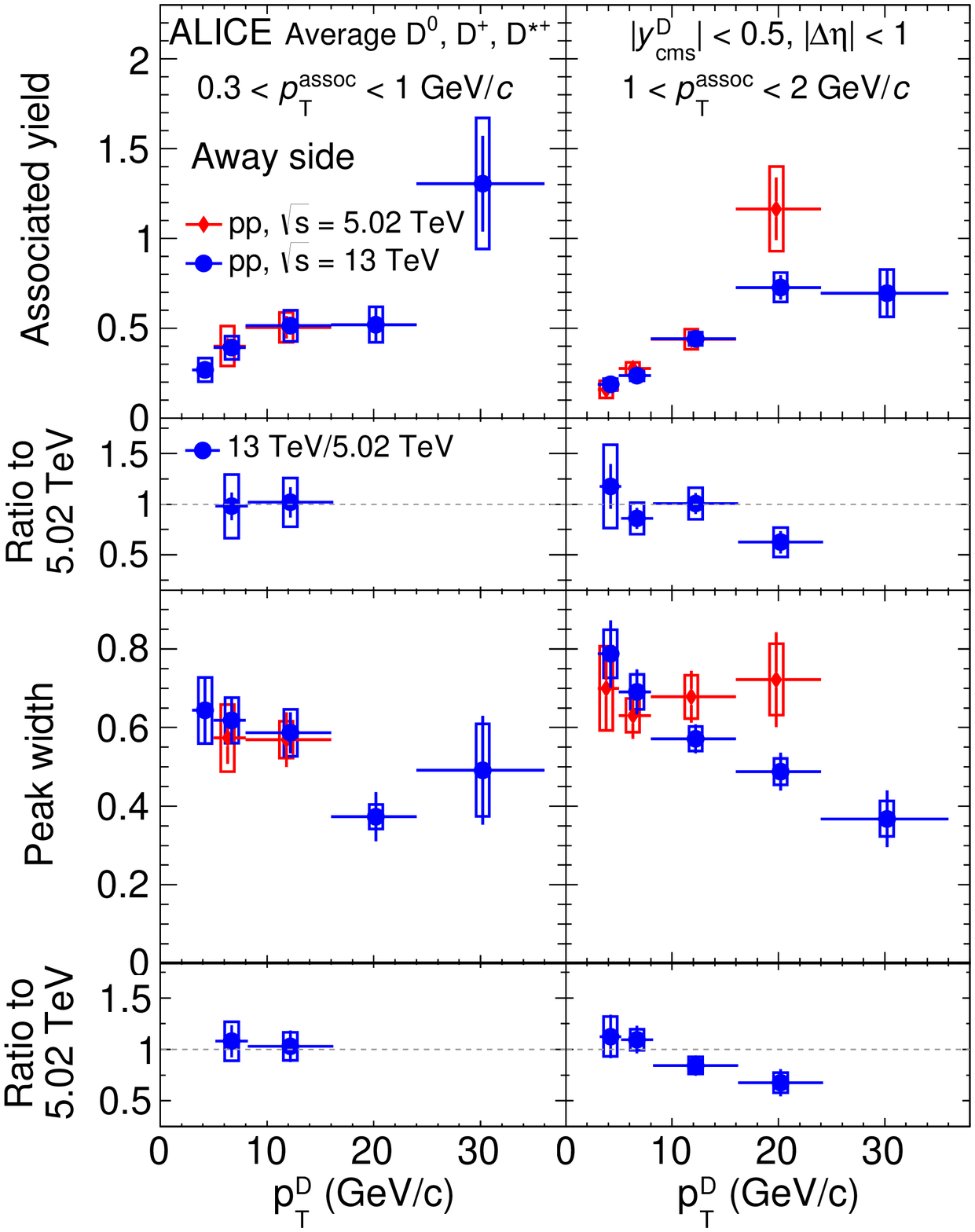}
	\caption {Near-side (left panel) and away-side (right panel) peak yields (first row) and widths (third row) obtained from a fit to the azimuthal correlation function after the baseline subtraction. The measurements are compared with ALICE results obtained in pp collisions at $\sqrt{s}$ = 5.02~TeV~\cite{Acharya:2019icl} and 7~TeV~\cite{ALICE:2016clc}, for 0.3 $<\ptass<$ 1 GeV/$c$, 1 $<\ptass<$ 2 GeV/$c$. Only near-side observables were computed in the $\sqrt{s}$ = 7~TeV analysis. The ratios to the $\sqrt{s} = 5.02$~TeV results are shown in the second and fourth rows for yields and widths, respectively.}%
	\label{fig:Comparisons_Energy_NS}%
\end{figure}

Figure~\ref{fig:Comparisons_Energy_DPhi} shows examples of correlation functions obtained from the analysis in pp collisions at a centre-of-mass energy of $\s$~=~13~TeV, compared with results previously reported by ALICE in pp collisions at $\s$~=~5.02~TeV~\cite{Acharya:2019icl} and $\s$~=~7~TeV~\cite{ALICE:2016clc} (the latter is available only for two kinematic ranges). The baseline value is closely related to the number of charged particles produced at midrapidity, and therefore has a strong dependence on $\s$, due to the increase of charged-particle production with increasing centre-of-mass energy~\cite{Adam:2015pza}. It was subtracted from the correlation functions in order to focus the comparison on the peak features. The shape of the distribution after the baseline subtraction and the properties of the correlation peaks at the three centre-of-mass energies agree within uncertainties. The analysis at $\s = 13$~TeV profits from a larger data sample, resulting in substantially smaller point-by-point statistical fluctuations (up to 50\% with respect to $\s = 5.02$~TeV results in the range $16<\ptD<24$~\GeVc), and leading to smaller uncertainties from the subtraction of the baseline.

More quantitative results are provided by the comparison of the near- and away-side peak yields and widths in pp collisions at different centre-of-mass energies, presented in Fig.~\ref{fig:Comparisons_Energy_NS}. The results for pp collisions at $\s = 7$~TeV were obtained after refitting the correlation functions measured in Ref.~\cite{ALICE:2016clc} with the improved function described in Eq.~\ref{eq:fit}, and evaluating the systematic uncertainties accordingly.
The near-side peak yield values obtained from the $\s = 13$~TeV data are compatible within the uncertainties with those at lower energies, exhibiting the same increasing trend of the yields with $\ptD$ in both the $\ptass$ intervals shown. An overall agreement is also observed between the $\s = 5.02$, 7, and 13~TeV results for the near-side widths. From the $\s = 13$~TeV results, an indication of a near-side peak narrowing for increasing $\ptD$ emerges, in particular for the $1 < \ptass < 2$~\GeVc range, that was not observed at lower energies because of the lower precision.
This peak narrowing could be originated by two simultaneous effects: (i) a more collimated angular pattern of the partons fragmented from charm quarks; (ii) an increased collinearity of charm and anti-charm quarks produced from gluon-splitting mechanism. Both effects are related to the increased boost, on average, of the fragmenting (splitting) parton when considering D-meson triggers with larger $\pt$.
An agreement within the uncertainties is also observed for the away-side peak results, and similar conclusions as those expressed for the near-side peak can be drawn. For the away-side observables only results at $\s = 5.02$~TeV are available for the comparison (and only for a restricted set of kinematic ranges), since the azimuthal-correlation functions for pp collisions at $\s = 7$~TeV were not precise enough to allow the characterisation of the away-side region, as discussed in Ref.~\cite{ALICE:2016clc}.

A similar comparison was performed using simulated pp collisions obtained with PYTHIA6 (with Perugia-2011 tune) and POWHEG+PYTHIA6~\cite{Alioli:2010xd,Nason:2004rx,Frixione:2007vw} event generators. A slight increase of the near-side yield values (5--10\% depending on the $\pt$ range) and a mild decrease of the away-side yield values (10--15\%) was observed when increasing the centre-of-mass energy from $\s = 5.02$ to 13~TeV, with small differences between the two generators. This could be ascribed to an increased contribution of NLO production processes (already included in the hard scattering in POWHEG+PYTHIA6, and accounted for during the parton-shower development in PYTHIA6), as well as to a harder charm-quark $\pt$ spectrum at larger centre-of-mass energies. These differences are within the overall precision of the data measurements. No visible energy dependence for both near- and away-side peak widths was found, for both generators, as observed in data.

\subsection{Results for different V0M multiplicity classes}
\label{sec:Result_vs_Centrality}
\begin{figure}[!t]
	\begin{center}
	\includegraphics[width=0.99\textwidth]{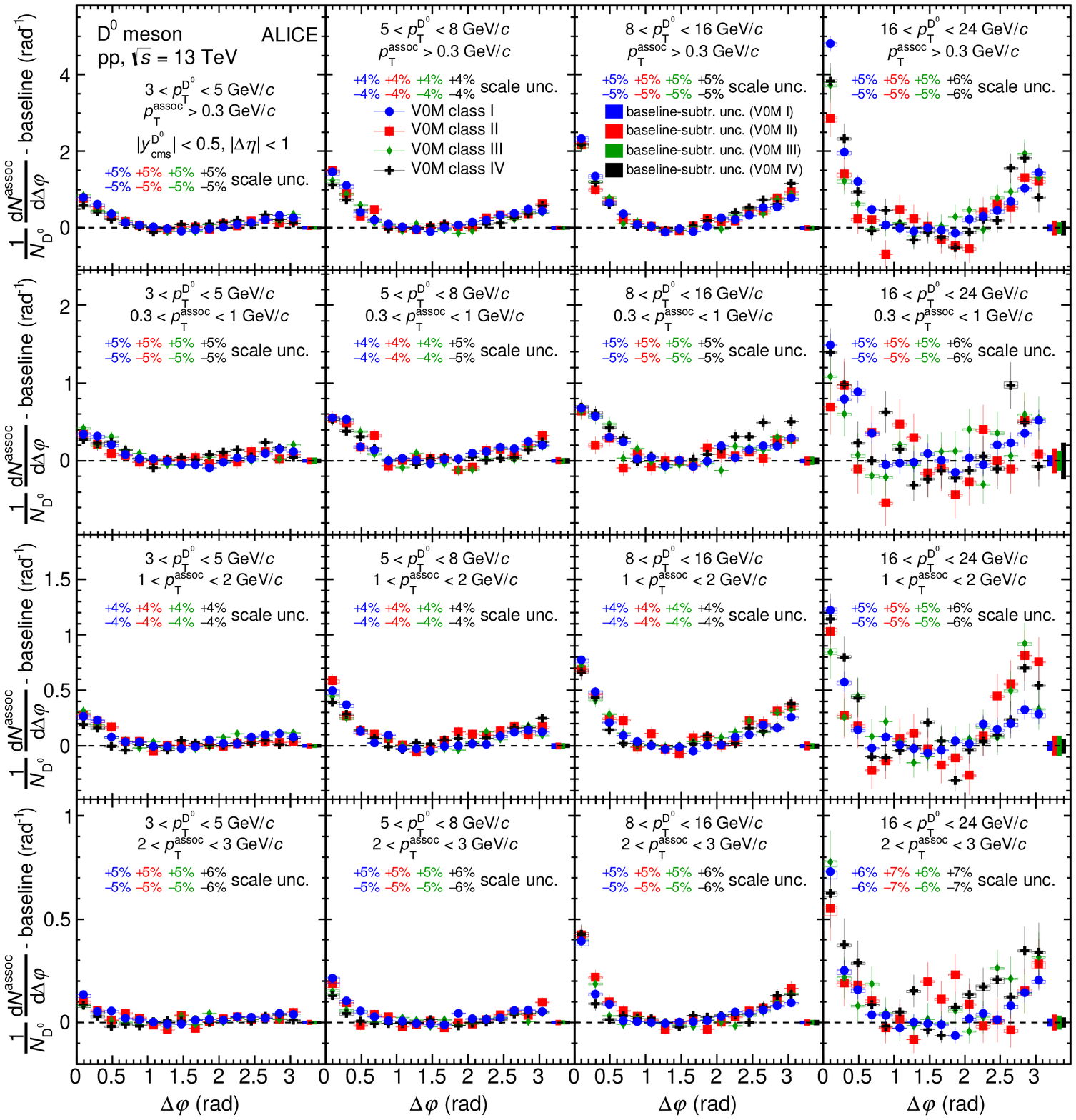}
	\end{center}
	\caption {Azimuthal-correlation functions of $\Dzero$ mesons with associated particles, after the subtraction of the baseline, in pp collisions at $\sqrt{s}$ = 13~TeV, for 3 $< \ptD <$ 5~\GeVc, 5 $< \ptD <$ 8~\GeVc, 8 $< \ptD <$ 16~\GeVc, and 16 $<\ptD<$ 24~\GeVc (from left to right) and $\ptass>$ 0.3~\GeVc, 0.3 $<\ptass<$ 1~\GeVc, 1 $<\ptass<$ 2~\GeVc, 2 $<\ptass<$ 3~\GeVc (from top to bottom) for different multiplicity classes estimated with V0M. The four multiplicity classes are shown with different marker styles. Statistical and $\Delta\varphi$-dependent systematic uncertainties are shown as vertical error bars and boxes, respectively, and $\Delta\varphi$-independent uncertainties are written as text. The uncertainties from the subtraction of the baseline are displayed as boxes at $\Delta\varphi>\pi$.}%
	\label{fig:Comparisons_Multiplicity_dPhy}%
\end{figure}

The azimuthal-correlation functions evaluated in the four classes of V0M multiplicity are compared in Fig.~\ref{fig:Comparisons_Multiplicity_dPhy}, for the four available $\Dzero$-meson $\pt$ ranges (one per column) and the four different $\ptass$ intervals (one per row). The baseline value largely increased from V0M multiplicity class IV towards V0M multiplicity class I, as expected due to the very different underlying-event activity, and it was subtracted from the correlation functions. This comparison suggests a similar shape and $\pt$ evolution of the azimuthal-correlation function, as well as consistent near- and away-side features for the four V0M multiplicity classes.

\begin{figure}[t]
	\centering
	\includegraphics[width=0.99\textwidth]{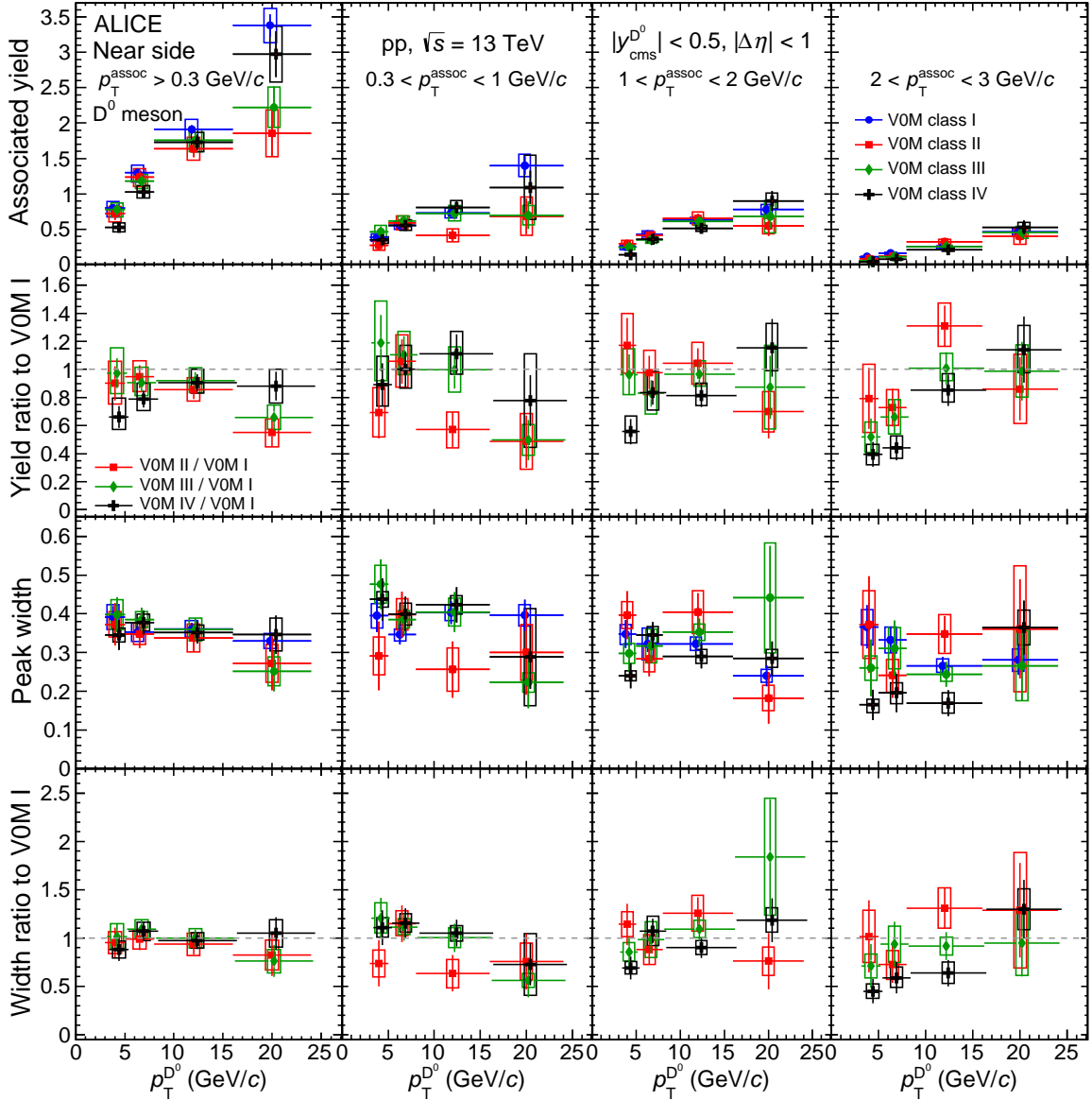}%
	\caption {Near-side associated peak yields (top row) and widths (third row) measured in pp collisions at $\sqrt{s}$~=~13~TeV, for the four V0M multiplicity classes, shown with different marker styles. The ratios of yield (width) values in each V0M class with respect to those in the V0M class I are shown in the second (fourth) row. Results are presented as a function of the $\Dzero$-meson $\pt$, for $\ptass>$ 0.3~\GeVc and the sub-ranges 0.3 $<\ptass<$ 1~\GeVc, 1 $<\ptass<$ 2~\GeVc, and 2 $<\ptass<$ 3~\GeVc (from left to right). Statistical and systematic uncertainties are shown as vertical error bars and boxes, respectively.}%
	\label{fig:Comparisons_Multiplicity_NS}%
\end{figure}

The near-side peak yield and width values obtained from the fit to the azimuthal correlation function in the four V0M multiplicity classes, for the same kinematic ranges, are shown in Fig.~\ref{fig:Comparisons_Multiplicity_NS} (first and third rows, respectively). Apart from a tension for low $\ptD$, 2 $<\ptass<$ 3~\GeVc, the yield measurements follow a similar increasing trend with $\ptD$.
Similar values are observed from the near-side peak widths.
These results indicate no significant modification of the charm fragmentation and hadronisation in collisions of varying charged-particle multiplicities.
The ratios of the yield and width results in V0M classes II, III, and IV over those in V0M class I, shown in the second and fourth rows of Fig.~\ref{fig:Comparisons_Multiplicity_NS}, respectively, also confirm this conclusion.

\begin{figure}[t]
    \begin{center}
    \includegraphics[width = 0.46\textwidth]{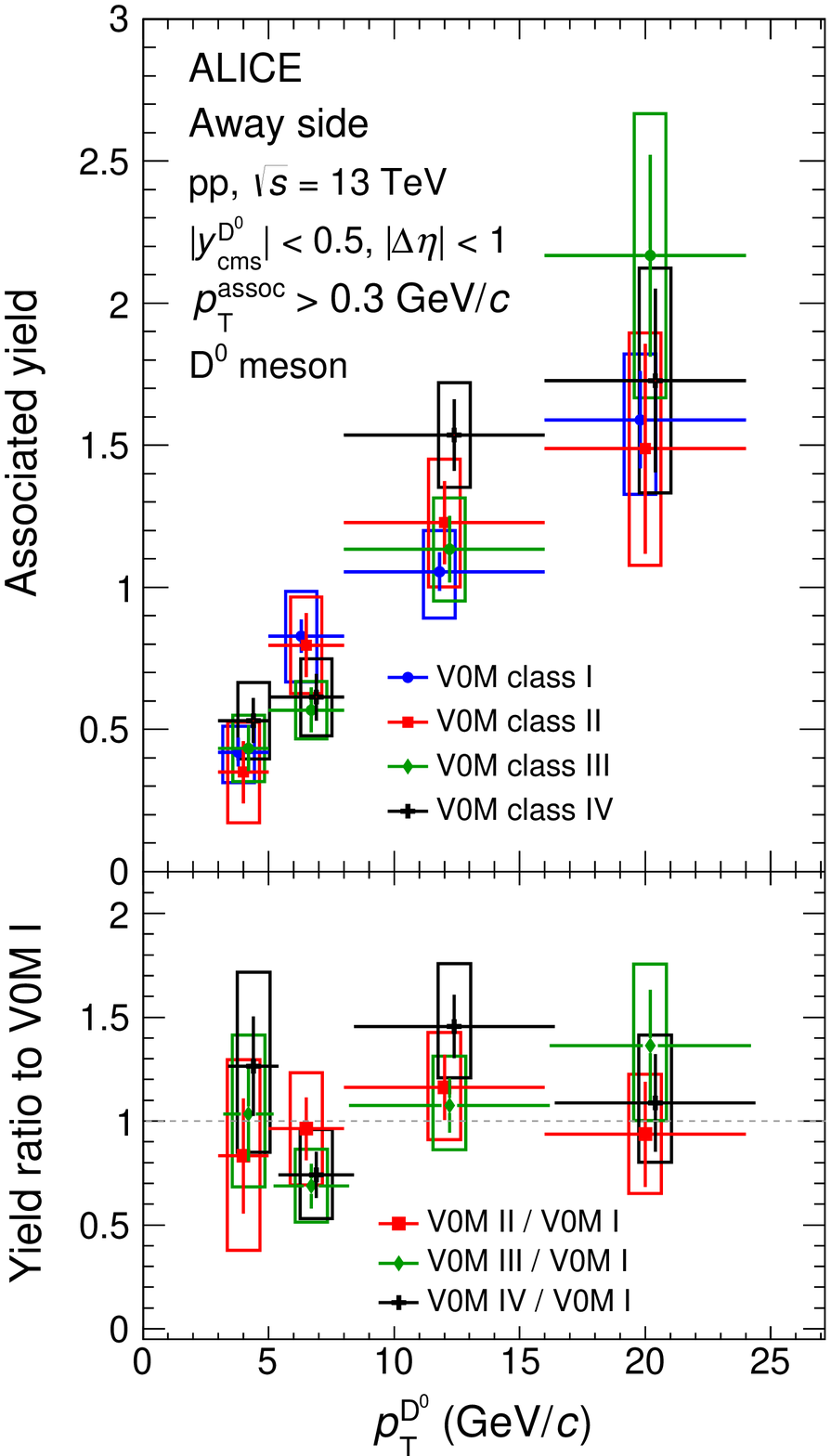}
    \includegraphics[width = 0.46\textwidth]{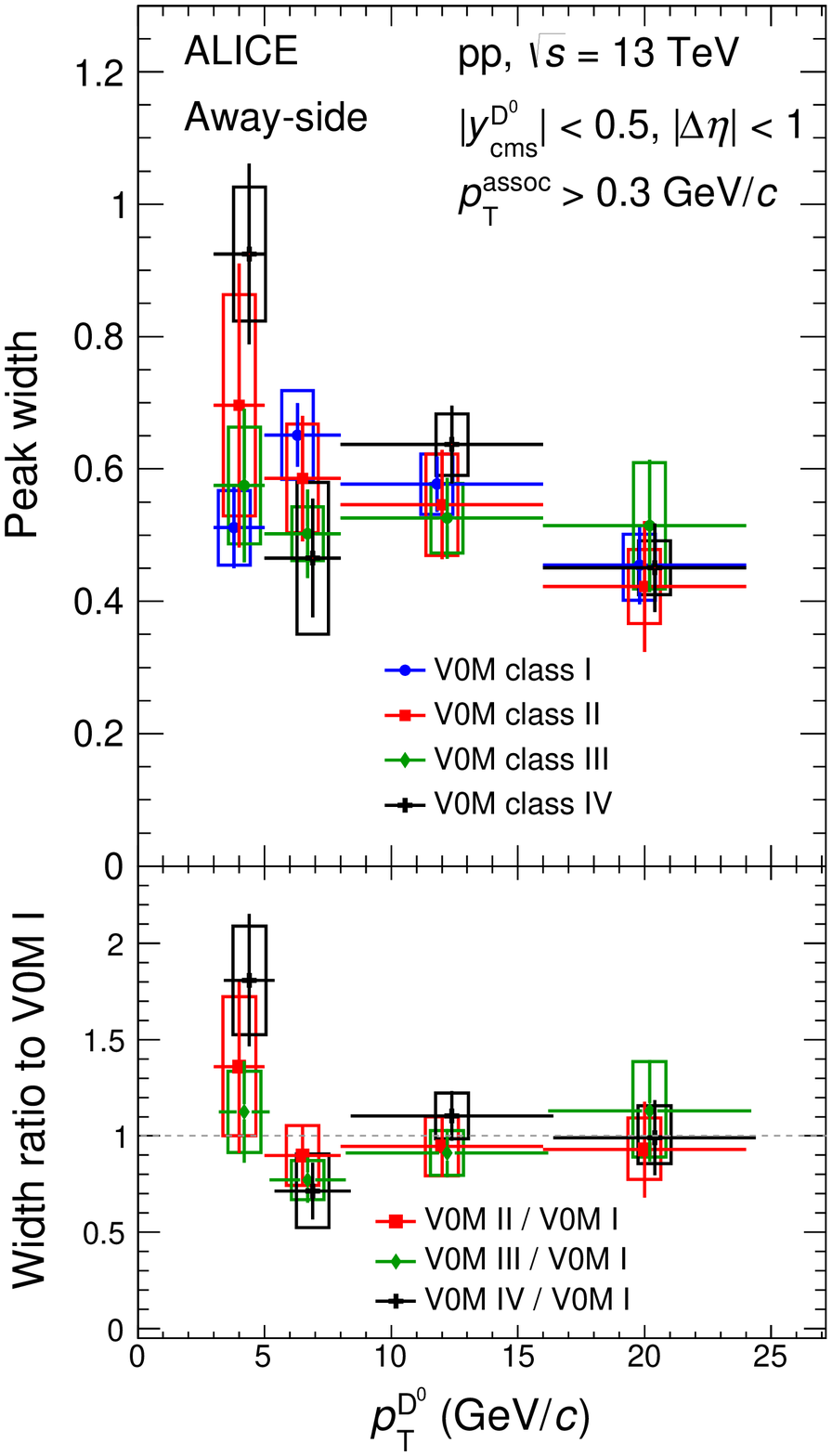}
    \end{center}
	\caption {Away-side associated peak yields (left) and widths (right) measured in pp collisions at $\sqrt{s}$~=~13~TeV, for the four V0M multiplicity classes, shown with different marker styles. The ratios of the observable values in each V0M class with respect to those in V0M class I are represented in the bottom insets.  Results are presented as a function of the $\Dzero$-meson $\pt$, for $\ptass>$ 0.3~\GeVc. Statistical and systematic uncertainties are shown as vertical error bars and boxes, respectively.}%
	\label{fig:Comparisons_Multiplicity_AS}%
\end{figure}

The evaluation of away-side peak yields and widths as a function of the event multiplicity were performed only in the integrated associated particle interval $\ptass > 0.3$~\GeVc, due to their large sensitivity to point-by-point statistical fluctuations. The away-side peak observable values for each of the four V0M multiplicity classes are shown in Fig.~\ref{fig:Comparisons_Multiplicity_AS}, together with the ratios of the values in the V0M classes II, III and IV over the values in the V0M class I. As observed for the near-side, the same increasing trend with $\Dzero$ $\pt$ is present for the yields, among the four V0M multiplicity classes. A hint of narrowing of the away-side peak, visible in the multiplicity-integrated results at $\s = 13$~TeV, can also be seen in the multiplicity-dependent results. The away-side yield and width values are fully consistent within the uncertainties among all four V0M classes.

Though with sizeable uncertainties, these measurements point towards consistency of the jet-induced correlation peak structure and shape in high- and low-multiplicity events, and thus contribute to confirm the assumptions done in the measurements of the elliptic-flow coefficient of charm particles in high-multiplicity pp collisions~\cite{CMS:2020qul,ATLAS:2019xqc}.

\subsection{Comparison of the ALICE results with model predictions}
\label{sec:Results_vs_MonteCarlo}
The near- and away-side peak yields measured in pp collisions at $\s = 13$~TeV and reported in Section~\ref{sec:Result_pp_vs_energies} were compared to predictions from several event generators. This allowed for verifying, for each model, whether its specific implementation of the processes leading from charm-quark production to final-state particles was adequate for describing the measured observables.
A detailed description of the models used for the comparison is provided in Ref.~\cite{Acharya:2019icl}. These models include PYTHIA8~\cite{Sjostrand:2007gs} with 4C~\cite{Corke:2010yf} tune, PYTHIA6~\cite{Sjostrand:2006za} with Perugia-2011 tune~\cite{Skands:2010ak}, POWHEG+PYTHIA8~\cite{Alioli:2010xd,Nason:2004rx,Frixione:2007vw} with hard-scattering matrix elements evaluated at NLO or at LO accuracy, HERWIG~7~\cite{Bahr:2008pv,Bellm:2015jjp}, and EPOS~3.117~\cite{Werner:2010aa,Drescher:2000ha}.

For each model, the average of the $\Dzero$, $\Dplus$, and $\Dstar$ azimuthal-correlation functions with charged particles was evaluated, using the same prescriptions applied for data analysis in terms of kinematic and particle-species selections.
The evaluation of the peak observables from the fit to the correlation distribution followed the same approach employed on data, except for the estimation of the baseline.
Since the statistical fluctuations in the transverse region are negligible for the models, the minimum of the azimuthal correlation function was directly considered as the baseline value. A systematic uncertainty on the peak observables was then assigned by performing an alternate fit, fixing the baseline as the weighted average of the two lowest points of the azimuthal-correlation function.

\begin{figure}[t]
	\centering
	{\includegraphics[width = 0.99\textwidth]{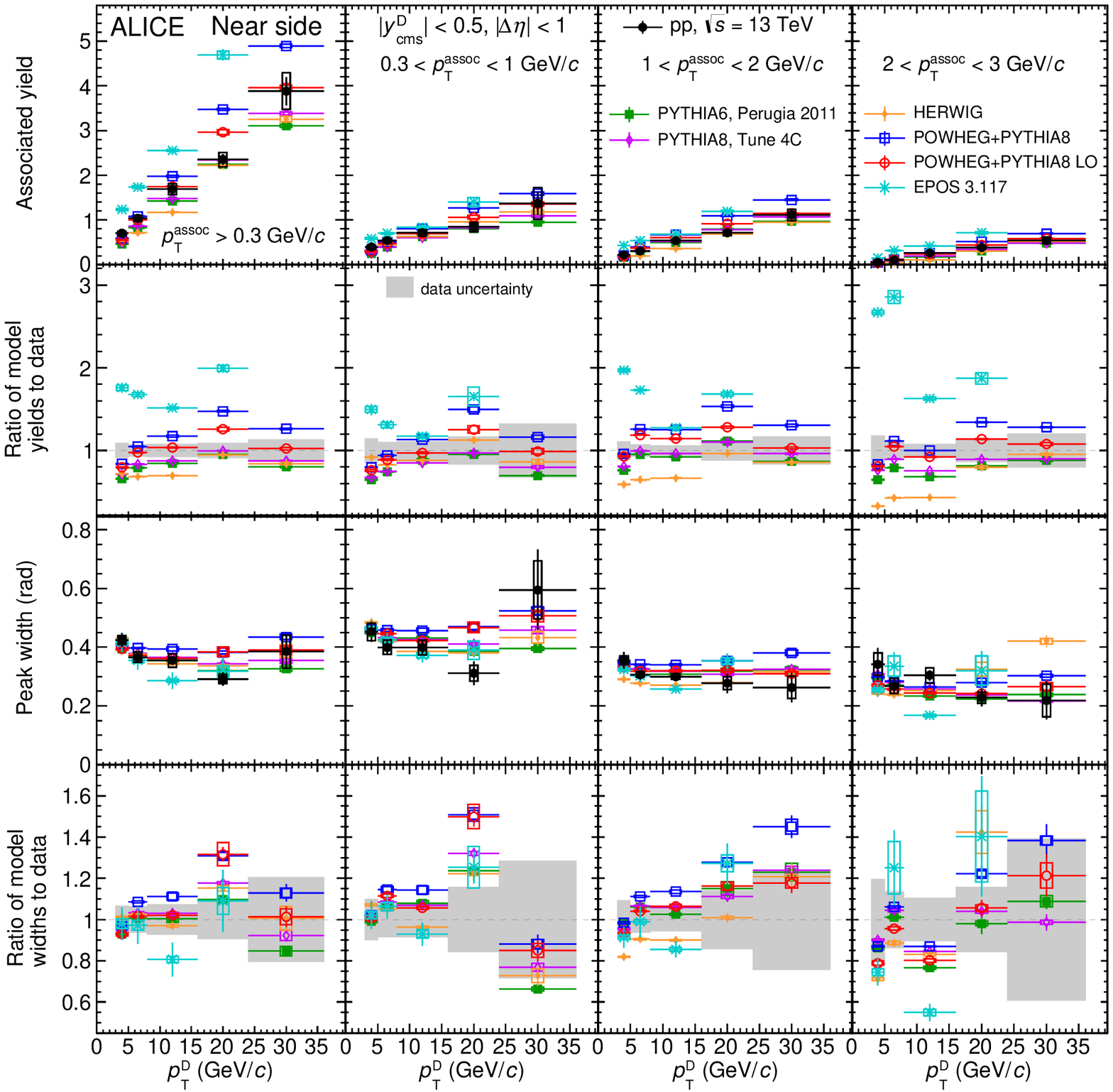} }%
	\caption {Near-side associated peak yields (top row) and widths (third row down) in pp collisions at $\sqrt{s}$ = 13~TeV, compared with predictions by PYTHIA, POWHEG+PYTHIA8, HERWIG, and EPOS 3 event generators with various configurations. Results are presented as a function of the D-meson $\pt$, for $\ptass>$ 0.3~\GeVc, $0.3 < \ptass < 1$~\GeVc, $1 < \ptass < 2$~\GeVc, and $2 < \ptass < 3$~\GeVc (from left to right). The ratios of model predictions to data measurements for yield (width) values are shown in the second (fourth) row down. In these rows, model statistical and systematic uncertainties are shown as vertical error bars and boxes, respectively, while data total uncertainties are displayed as a solid band.}%
	\label{fig:Comparisons_Models_NS}%
\end{figure}

The near-side peak observable trends for both models and data are illustrated in Fig.~\ref{fig:Comparisons_Models_NS} as a function of the trigger D-meson $\pt$, in the $\ptass > 0.3$~\GeVc interval and in the other three kinematic sub-ranges. The first and third rows show the yield and the width values, while the second and the fourth show the ratios of model predictions with respect to data. In these ratio panels, model statistical and systematic uncertainties are shown as error bars and boxes, respectively, while data statistical and systematic uncertainties are summed in quadrature, and the resulting uncertainty is represented as a solid grey band.
The increasing trend of the near-side yield with the trigger particle $\pt$ seen in data is obtained by all the MC predictions, but with different strengths. A hierarchy can be observed for the yield values, with EPOS systematically providing the largest yields, followed by POWHEG+PYTHIA8 NLO, POWHEG+PYTHIA8 LO, and then by PYTHIA6 and PYTHIA8. HERWIG predicts the lowest yields for $\ptD < 8$~\GeVc and $\ptass > 1$~\GeVc. Its predicted yield values for the other kinematic ranges, instead, are generally in between PYTHIA and POWHEG+PYTHIA8.
The best description of the measurements is provided by the POWHEG+PYTHIA8 and by PYTHIA generators, with POWHEG+PYTHIA8 (both NLO and LO) generally performing better at lower $\ptD$ and PYTHIA (both versions) at higher $\ptD$. A slight difference is observed between the NLO and LO implementations of POWHEG+PYTHIA8, with the former providing larger yields (by 5 to 15\%, increasing with the D-meson $\pt)$, and overall providing a better description of the data in the lower $\ptD$ region, while the latter has a better agreement with data above 8~\GeVc. These differences can be understood in terms of the different relative contribution of the NLO production mechanisms, as already discussed in Ref.~\cite{Acharya:2019icl}.
HERWIG predictions tend to underestimate the value of the near-side yield in the kinematic region $\ptD < 8$~\GeVc and $\ptass > 1$~\GeVc, while for the other kinematic regions the predictions are compatible with the data. EPOS predictions (not available for the range $24 < \pt < 36$~\GeVc) overestimate the near-side yield measurements by a factor of about 2 through all the studied kinematic ranges.
Generally, smaller differences are obtained for the widths of the various models with respect to those observed for the yields. POWHEG+PYTHIA8 NLO 
predicts the broadest near-side peaks. The near-side width data measurements hint towards a slight sharpening of the near-side peak width with increasing $\ptD$, while most of the models describe the width as nearly flat. However, all models are able to reproduce the measured width within the uncertainties.

\begin{figure}[t]
	\centering
	{\includegraphics[width = 0.99\textwidth]{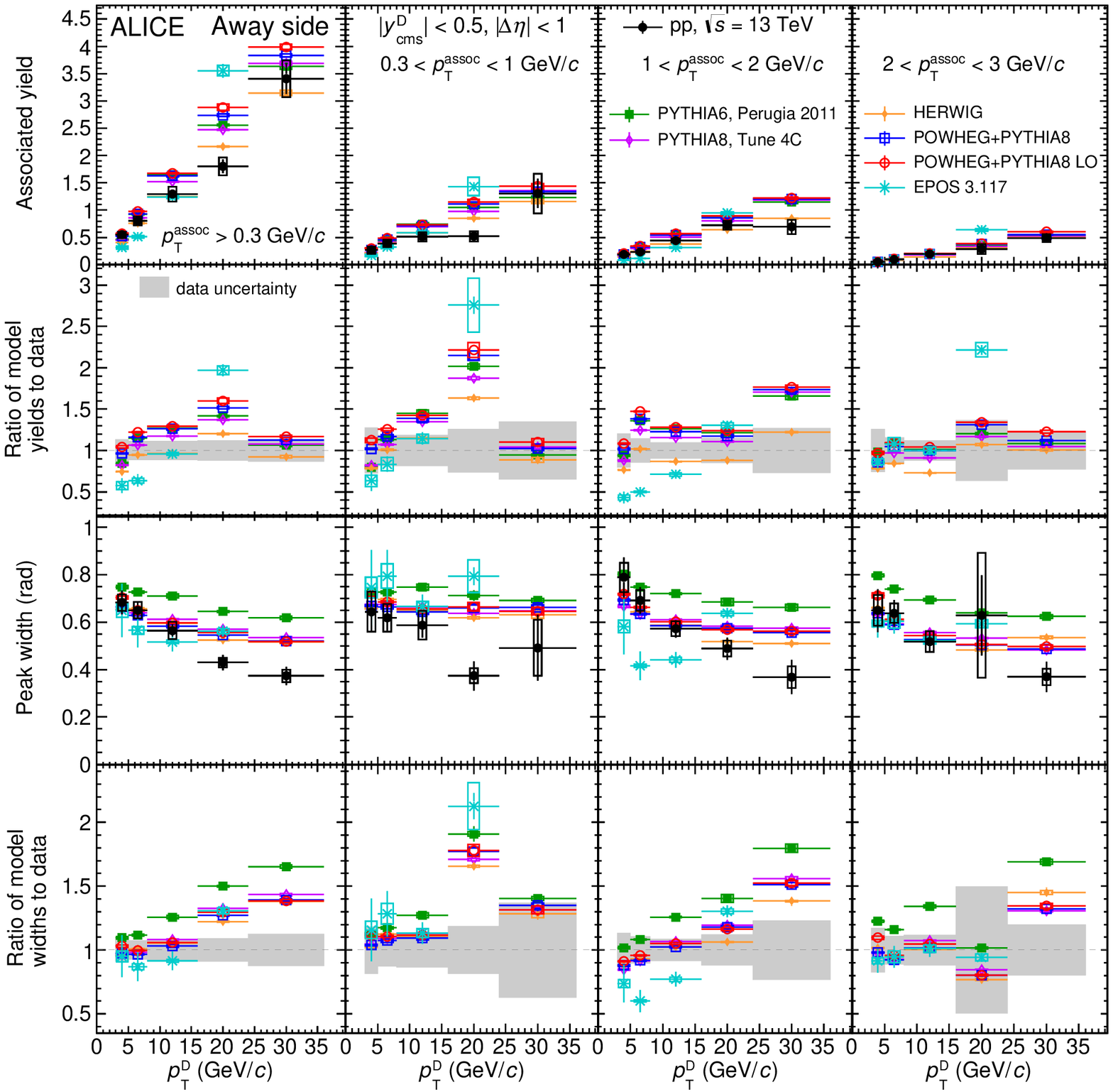} }%
	\caption {Away-side associated peak yields (top row) and widths (third row down) in pp collisions at $\sqrt{s}$ = 13~TeV, compared with predictions by PYTHIA, POWHEG+PYTHIA8, HERWIG, and EPOS 3 event generators with various configurations. Results are presented as a function of the D-meson $\pt$, for $\ptass>$ 0.3~\GeVc, $0.3 < \ptass < 1$~\GeVc, $1 < \ptass < 2$~\GeVc, and $2 < \ptass < 3$~\GeVc (from left to right). The ratios of model predictions to data measurements for yield (width) values are shown in the second (fourth) row down. In these rows, model statistical and systematic uncertainties are shown as vertical error bars and boxes, respectively, while data total uncertainties are displayed as a solid band.}%
	\label{fig:Comparisons_Models_AS}%
\end{figure}

A similar comparison for the away-side peak is shown in Fig.~\ref{fig:Comparisons_Models_AS}.
POWHEG+PYTHIA8 NLO and LO implementations provide the highest away-side yields, with the LO implementation generally 5\% above the NLO one, possibly due to an increased amount of back-to-back production processes. PYTHIA6 and PYTHIA8 generally provide slightly lower away-side yield values than POWHEG+PYTHIA8 NLO and LO expectations, with PYTHIA8 tending to be on the lower side compared to PYTHIA6. HERWIG predicts values lower than all the other models (about 20\% lower than POWHEG+PYTHIA8 NLO yields). EPOS predicts a stronger increasing trend of the away-side yields with $\ptD$ than the other generators.
For this observable, data uncertainties are not small enough to be sensitive to all the differences highlighted above. The best agreement with data is provided by PYTHIA6, PYTHIA8, and HERWIG. The POWHEG+PYTHIA8 predicted yields, for both implementations, are on the higher side compared to data, in particular for lower associated-particle $\pt$. EPOS predictions tend to underestimate the yield for $\ptD < 5$~\GeVc, while for $16 < \ptD < 24$~\GeVc its away-side yield predictions are higher than the measured values.
The narrowing of the away-side peak with increasing $\ptD$ is clearly evident in the third row of Fig.~\ref{fig:Comparisons_Models_AS}, both for model predictions and data, except for $0.3 < \ptass < 1$~\GeVc, where all models predict a rather flat trend. Only EPOS has a slightly different behaviour compared to the other models, showing an approximately flat $\ptD$ trend of the away-side width over the full $\ptass$ range, albeit with quite large model uncertainties. In terms of absolute values, PYTHIA6 away-side width expectations are systematically higher than all the other models, by an overall 20$\%$, with increasing differences for increasing $\ptass$, and also tend to overestimate the measured width values. All the other models predict similar values, consistent with data except for $\ptD > 16$~\GeVc, where they slightly overestimate the data measurements.

\section{Parton-level studies with PYTHIA8 and POWHEG+PYTHIA8}
\label{sec:Detailed_sim}

From the comparative studies discussed in Section~\ref{sec:Results_vs_MonteCarlo}, PYTHIA8 and POWHEG+PYTHIA8 provide, overall, the most accurate description of the near- and away-side correlation peak features. A more detailed investigation was performed using these models to expose how different stages of the event generation that determine the formation of the final-state particles generally influence the development of the features of the correlation peak and the azimuthal-correlation function.

At large momentum or short distances, e.g.~the hard-parton scattering leading to heavy quark production, QCD is asymptotically free. It implies that the coupling constant is small, so a perturbative approach can be employed. Before hard scattering takes place, partons from the incident beam protons can radiate gluons in the so-called initial-state radiation (ISR) process. Similarly, outgoing partons from the hard scatterings can produce a shower of softer particles via a final-state radiation (FSR) process. Since hadrons are composite objects, more than one distinct hard-parton interaction can occur in a pp collision, and proton remnants can also scatter again on each other. Such processes are called multi-parton interactions (MPI), and are responsible for the production of a large fraction of the particles uncorrelated with the D-meson candidate trigger, giving a substantial contribution to the underlying event (UE) observed in the collision final state.
Additionally, as detailed in Ref.~\cite{Adam:2015ota}, in the MPI implementation used in PYTHIA8 (which also drives the MPI process in POWHEG+PYTHIA8 simulations) charm-quark production can occur not only from the first (hardest) hard scattering, but also from hard processes in the various MPI occurring in the collisions, ordered with decreasing hardness.
There is also some correlation between FSR+ISR and MPI processes, since initial- and final-state radiations are generated from all the parton interactions occurring in the collision, and are thus enhanced in presence of MPI, further increasing the collision multiplicity.

In the following, PYTHIA8 and POWHEG+PYTHIA8 predictions, for standard simulations and for events generated after deactivating the previously mentioned parton-level contributions (FSR+ISR, and FSR+ISR+MPI), are compared. The latter case, in particular, provides a detailed view of the correlation function from the hard-scattering outgoing partons, though hadronisation and decays are still present and partially modify the original, parton-level, correlation shape.
The same procedure was performed for the predictions of near- and away-side peak yields, widths, and baseline value, which are then compared with the data, and of the $\beta$ parameter of the near-side peak.

\begin{figure}[t]
	\centering
	{\includegraphics[width=0.99\textwidth]{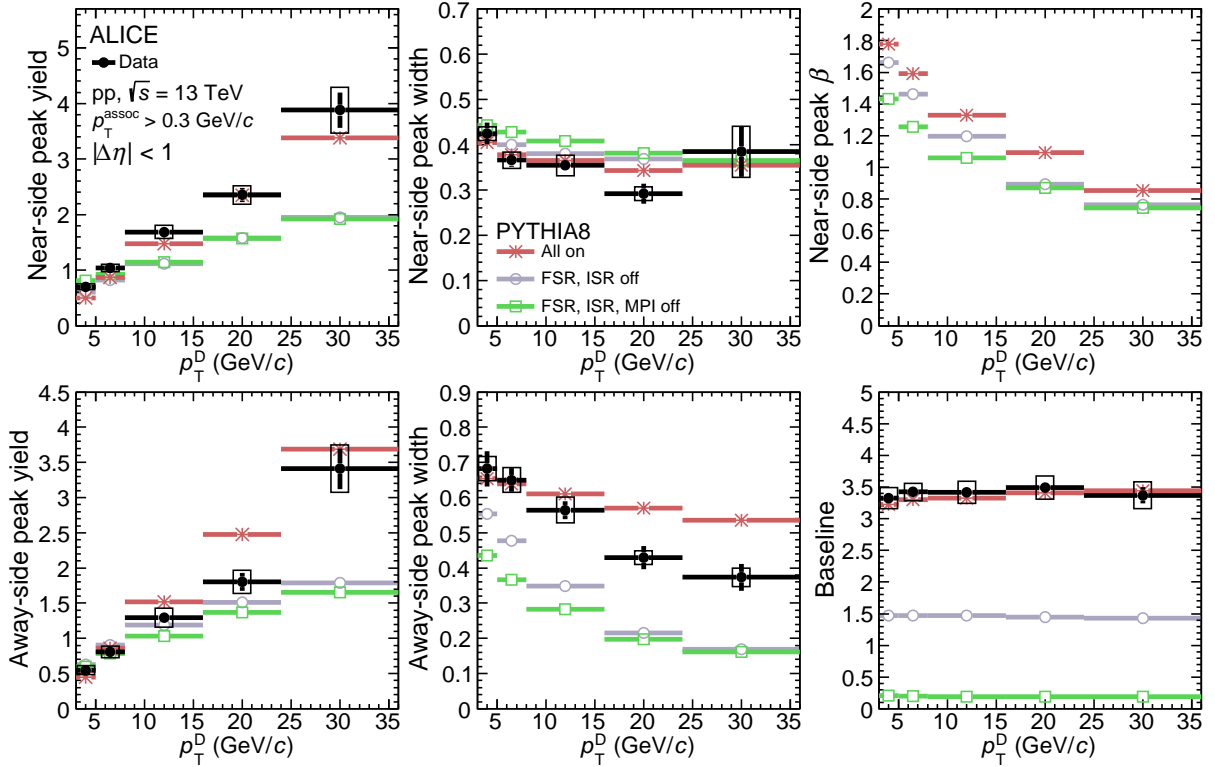} }%
	\caption {Near- and away-side peak yields (first column), widths (second column), near-side peak $\beta$ parameter and baseline (third column) from fits to the D-meson and charged particle azimuthal-correlation function, from PYTHIA8 simulations obtained with different parton-level contributions. The predictions are obtained for multiplicity-integrated \pp collisions at $\s = 13$ \TeV, as a function of the D-meson $\pt$, for $\ptass > 0.3$ \GeVc, and compared with ALICE data.}%
	\label{fig:partonpy}%
\end{figure}

Figure~\ref{fig:partonpy} shows the observables extracted from the fit to the azimuthal correlation function for different parton-level contributions from PYTHIA8 event generator, compared with the data, for $\ptass > 0.3$~\GeVc and as a function of $\ptD$.  In the first column, at high $\ptD$ the near-side and away-side yields show no relevant contribution of MPI, while FSR and ISR processes contribute to an increase of both peak yields, as expected, since for high-momentum partons they lead to additional production of collinear particles.
Even at high $\ptD$, with all three processes switched off, PYTHIA8 predicts peak yields amounting to roughly half of the measured yields.
In the lowest $\ptD$ interval, instead, FSR, ISR, and MPI lead to a decrease of the peak yields.
In the second column, the near-side width shows no significant modification for the various configurations, apart from a slight increase of the width observed when switching off MPI. 
These insights point towards a relevant role of hadronisation, which remains in place also in the absence of FSR, ISR, and MPI processes, in shaping the near-side correlation peak.
As expected, the away-side peak is wider than the near-side peak because of a combined contribution of parton-level effects. When FSR and ISR are turned off, the peak width is substantially decreased. This could be explained by the lack of radiation (in particular hard gluon emissions) decreasing the deflection angle of the recoil jet.
In the third column, a mild dependence is observed for the near-side $\beta$ parameter. In the simulations, the available sample is much larger than in data, so the $\beta$ parameter could be left free in the fit function, and its value is compared among the different model configurations. A very small contribution of parton-level processes to the baseline comes directly from the hard scattering (and subsequent hadronisation and decays), as expected since PYTHIA8 is a LO generator. All the other processes (ISR, FSR, and MPI) contribute to the baseline, with the MPI roughly equivalent to the sum of ISR and FSR. This is expected since MPI mostly affect the underlying event, whose particles point to directions largely unrelated to the trigger D-meson one. The further increase of the baseline when activating ISR and FSR processes is due to the fact that, as mentioned above, they act on all parton scatterings, including those occurring in MPI.

\begin{figure}[t]
	\centering
	{\includegraphics[width=0.99\textwidth]{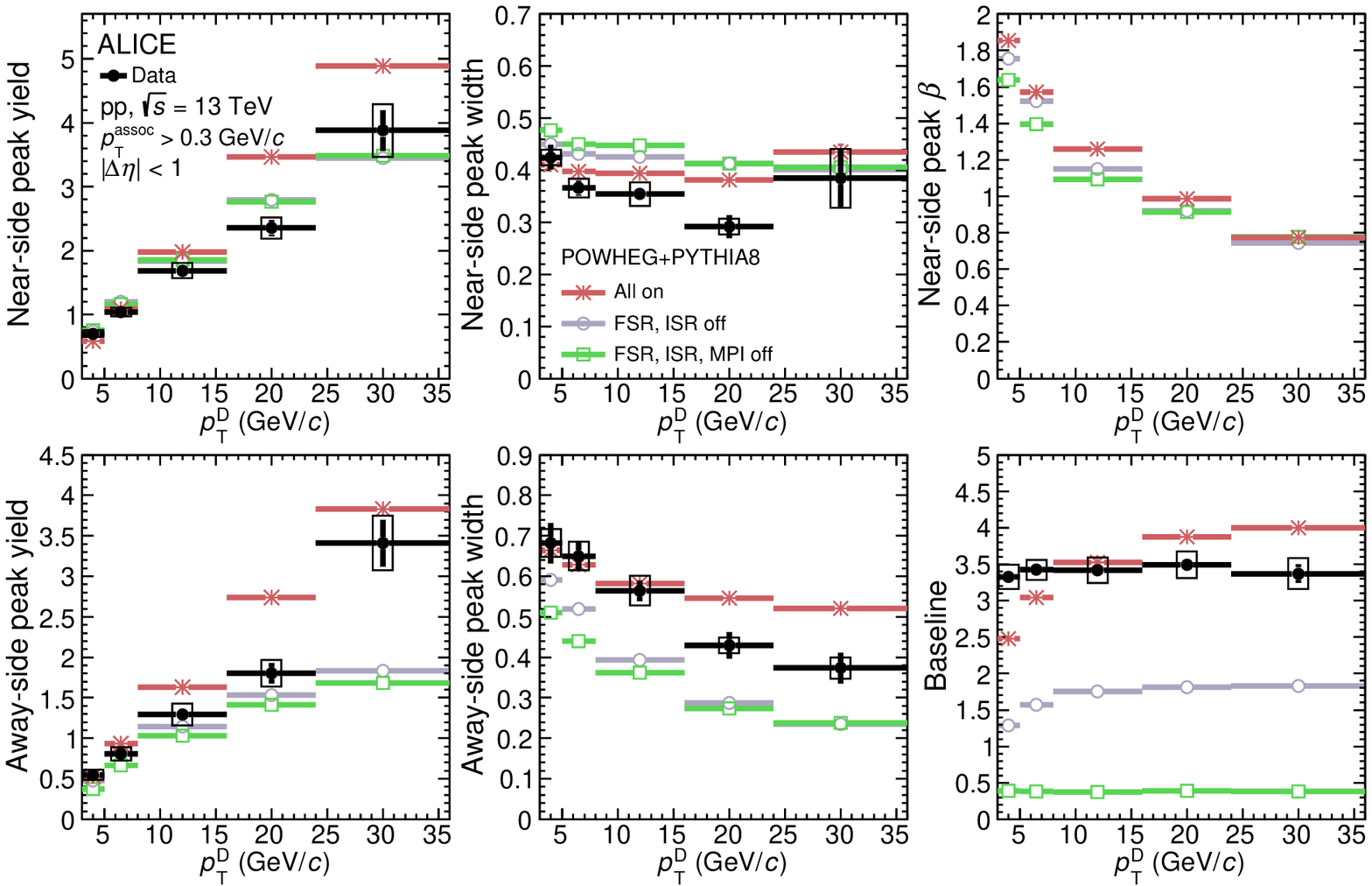} }%
	\caption{Near- and away-side peak yields (first column), widths (second column), near-side peak $\beta$ parameter and baseline (third column) from fits to the D-meson and charged particle azimuthal-correlation function, from POWHEG+PYTHIA8 simulations obtained with different parton-level contributions. The predictions are obtained for multiplicity-integrated in \pp collisions at $\s = 13$ \TeV, as a function of the D-meson $\pt$, for $\ptass > 0.3$ \GeVc, and compared with ALICE data.}%
	\label{fig:partonpow}%
\end{figure}

Figure~\ref{fig:partonpow} shows the same set of observables as shown in Fig.~\ref{fig:partonpy} and provides the same comparison to data, but with predictions obtained from POWHEG+PYTHIA8. Compared to Fig.~\ref{fig:partonpy}, a larger near-side peak yield is already obtained when ISR, FSR, and MPI processes are switched off. This can be explained with the inclusion of NLO processes directly in the hard scattering in POWHEG, rather than reproducing their effect in the parton shower as in PYTHIA8. Also in this case, MPI does not contribute to the peak yield. A similar behaviour as that of Fig.~\ref{fig:partonpy} is observed for the near-side peak width. The near-side peak $\beta$ parameter value shows a decreasing influence of ISR, FSR, and MPI processes for increasing $\ptD$. For the away-side peak, no major differences with respect to PYTHIA8 are found for the yield contributions, while a slightly lower influence of ISR and FSR is obtained for the widths, which can also be understood with the above consideration. For the baseline, a higher value with respect to PYTHIA8 is obtained when switching off all the parton-level processes, consistent with the non back-to-back topology of NLO processes (in particular, flavour excitation with a nearly flat contribution in $\Delta\varphi$~\cite{Mangano:1991jk}). The FSR and ISR contributions to the baseline increase with $\ptD$, in contrast to PYTHIA8, and partially drive the rising $\ptD$ trend of the baseline observed for the full simulation.

\begin{figure}[t]
	\centering
	\begin{minipage}{.49\textwidth}
		\centering
		\includegraphics[width=\textwidth]{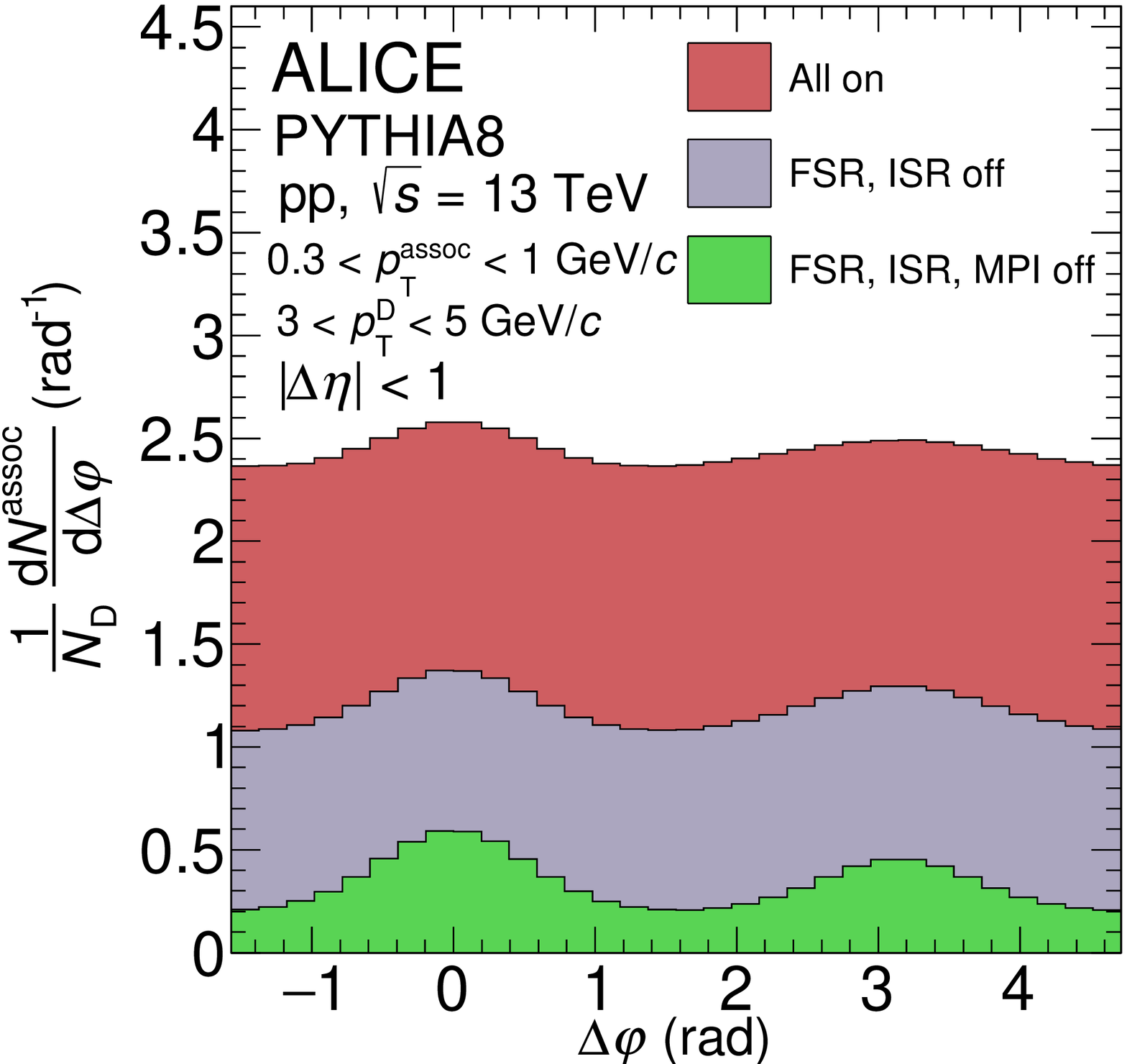}
	\end{minipage}%
	\hfill
	\begin{minipage}{.49\textwidth}
		\centering
		\includegraphics[width=\textwidth]{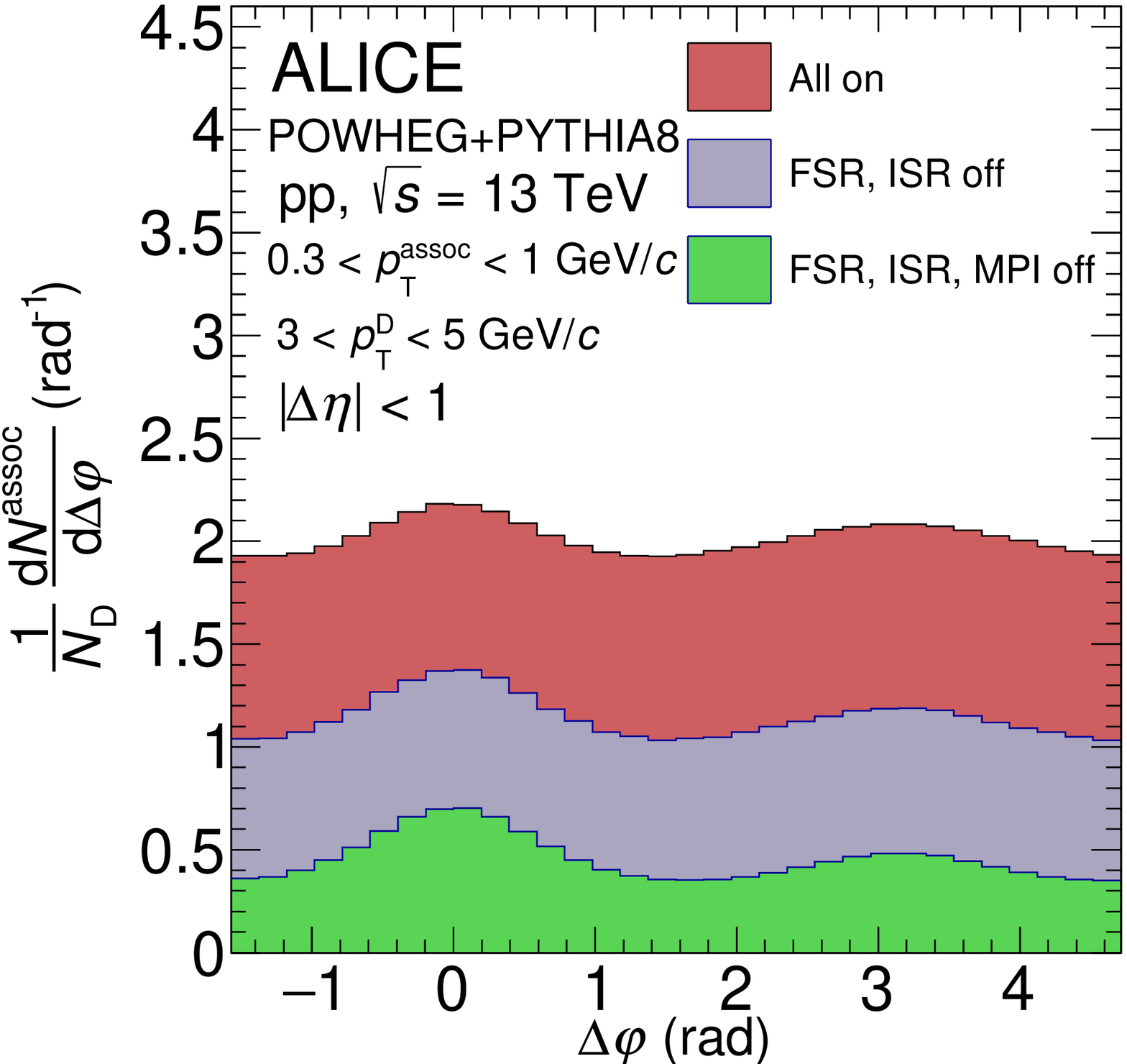}
	\end{minipage}
	\caption{Azimuthal correlation function of D mesons with charged particles with different parton-level contributions from PYTHIA8 (left panel) and POWHEG+PYTHIA8 simulations (right panel) in pp collisions at $\s = 13$~\TeV, for $0.3 <\ptass< 1$~\GeVc and $3 <\ptD< 5$~\GeVc.}
	\label{fig:sim_Dh_1}
\end{figure}

\begin{figure}[t]
	\centering
	\begin{minipage}{.49\textwidth}
		\centering
		\includegraphics[width=\textwidth]{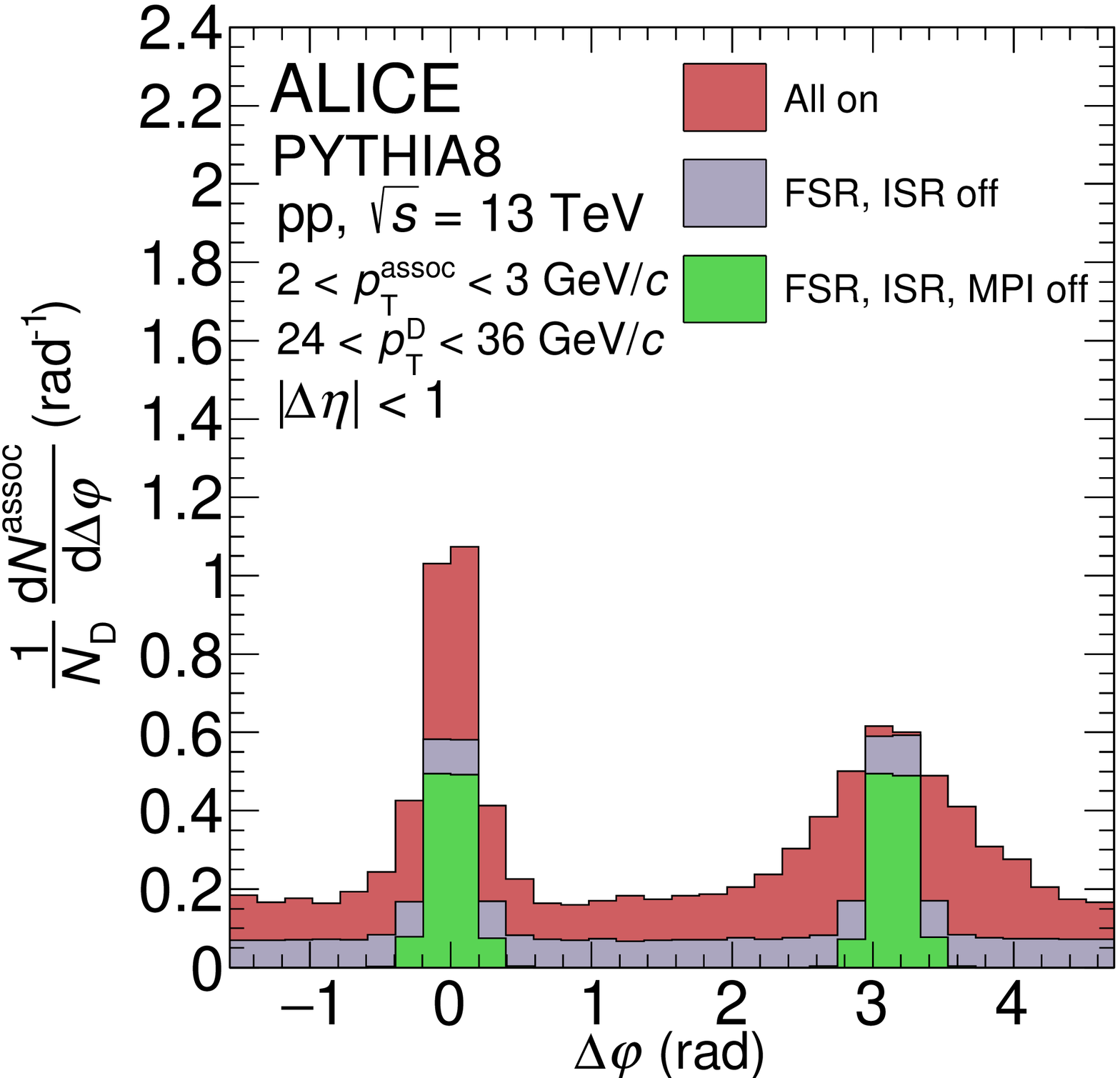}
	\end{minipage}%
	\hfill
	\begin{minipage}{.49\textwidth}
		\centering
		\includegraphics[width=\textwidth]{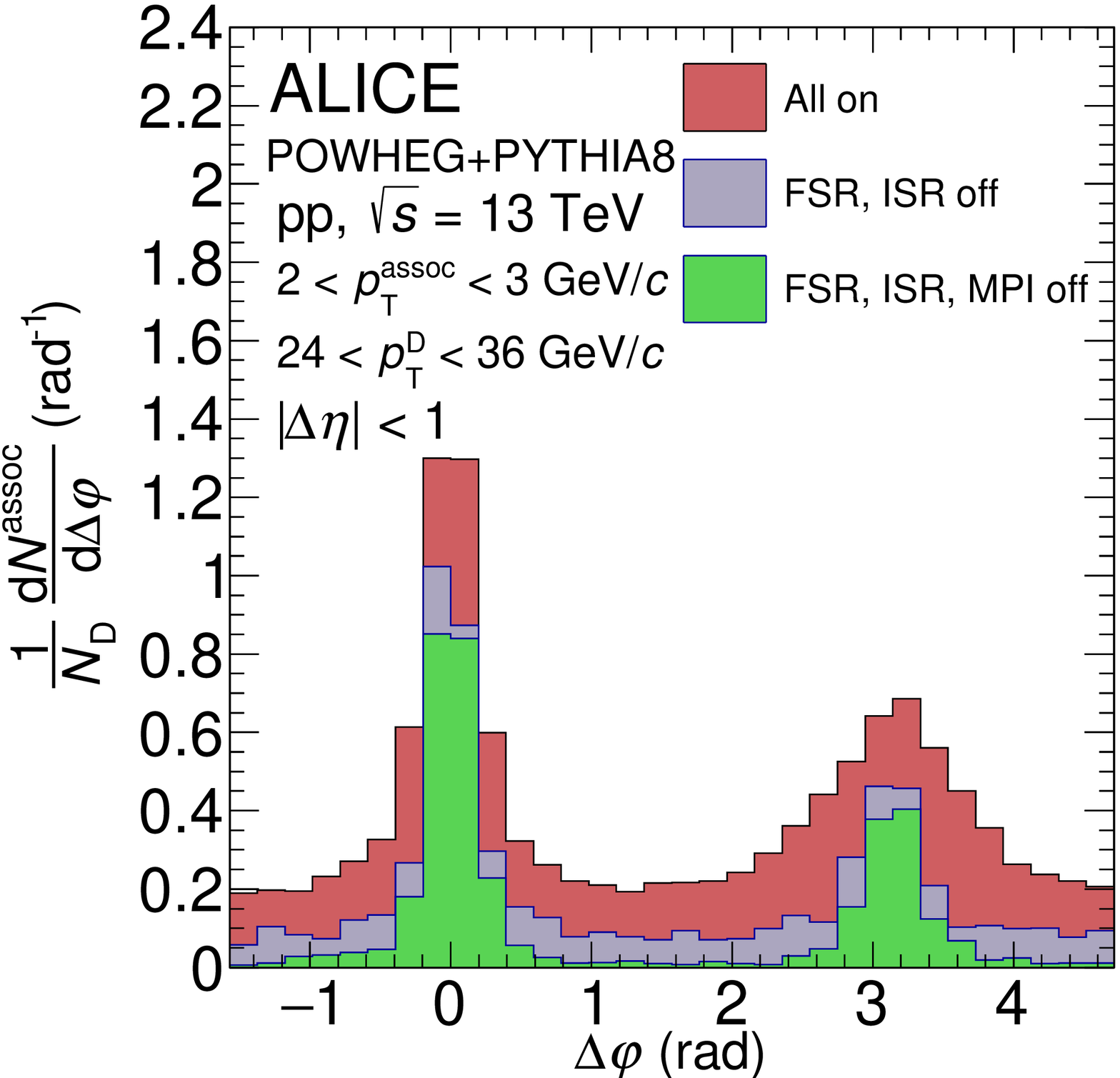}
	\end{minipage}
	\caption{Azimuthal correlation function of D mesons with charged particles with different parton-level contributions from PYTHIA8 (left panel) and POWHEG+PYTHIA8 simulations (right panel) in pp collisions at $\s = 13$~TeV, for $2 <\ptass< 3$~\GeVc and $24 <\ptD< 36$~\GeVc.}
	\label{fig:sim_Dh_2}
\end{figure}

The comparison of the contributions of the various processes to the correlation function helps in understanding better the source of the difference between PYTHIA8 and POWHEG+PYTHIA8 simulations observed in Section~\ref{sec:Results_vs_MonteCarlo}. Figure~\ref{fig:sim_Dh_1} displays the azimuthal correlation function of D mesons with charged particles obtained from PYTHIA8 and POWHEG+PYTHIA8 simulations sequentially deactivating the different parton-level contributions in pp collisions at $\s = 13$~\TeV for $0.3 <\ptass< 1$~\GeVc and $3 <\ptD< 5$~\GeVc.
Figure~\ref{fig:sim_Dh_2} shows the same quantities for a higher momentum range, i.e.~$2 <\ptass< 3$~\GeVc and $24 <\ptD< 36$~\GeVc.

Most of the features expressed above are seen in a more qualitative but clearer way in these figures. In particular, the larger baseline in the case without FSR, ISR, and MPI for POWHEG+PYTHIA8 is clearly visible for the low-$\pt$ range.
In general, for both simulations the relevant ISR, FSR, and MPI contributions to the baseline are obtained over the whole $\Delta \varphi$ range when focusing in the low-momentum region, while most of the off-peak contribution disappears when considering higher-$\pt$ trigger and associated particles. In the high-$\pt$ region, the difference in the contributions to the peaks between the generators is evident: in particular, for POWHEG+PYTHIA8 the near-side peak yield is already larger for the case without ISR, FSR, and MPI processes. Also, at high $\pt$ two sharp peaks appear for PYTHIA8, when parton showering and MPI effects are turned off, while for POWHEG+PYTHIA8 wider peaks emerge already from the hard-scattering, due to higher-order charm-production processes. The addition of the parton showering processes in PYTHIA8 allows to reconcile most of the differences in the correlation peak shape, in particular for the widths, while some residual differences remain present for the yields for the full simulation, as discussed in Section~\ref{sec:Results_vs_MonteCarlo}.

\section{Summary}
\label{sec:Summary}

A study of the azimuthal correlation function of D mesons with charged particles, measured in pp collisions at $\s = 13$ TeV with the ALICE detector, was presented. The pattern of the correlation function and the features of near- and away-side correlation peaks, extracted via a fit to the correlation function, were characterised in five D-meson momentum ranges, from 3 to 36 \GeVc, and for the associated-particle range $\ptass > 0.3$ \GeVc and the three sub-ranges $0.3 < \ptass < 1$~\GeVc, $1 <\ptass< 2$~\GeVc, and $2 <\ptass < 3$~\GeVc.

The measurement precision is significantly improved compared to previous ALICE results in pp collisions at $\s = 7$ TeV~\cite{ALICE:2016clc} and $\s = 5.02$ TeV~\cite{Acharya:2019icl}. The correlation function shape, as well as the peak yields
and widths, are compatible within uncertainties with those lower-energy measurements, confirming the expectation from PYTHIA8 and POWHEG+PYTHIA8 generators of little dependence on the collision energies.

The possible evolution of the correlation function with the event multiplicity was probed by performing the analysis in four multiplicity ranges, measured with the V0M estimator, profiting from a dedicated high-multiplicity trigger provided by the V0 detector. Though the uncertainties do not allow for firm conclusions, a strong variation of the correlation function with multiplicity is excluded, suggesting that when the charm quarks hadronise into $\Dzero$ mesons, the charm-quark fragmentation and hadronisation processes are not particularly sensitive to the event multiplicity. The overall compatibility of the correlation-peak features for different event multiplicities tends to support the scenario that, independently of the number of charm quarks produced in the collision, they undergo similar fragmentation and hadronisation into $\Dzero$ mesons. Such a scenario indicates an increased, but independent, charm production from MPI in high-multiplicity pp collisions, and is one of the mechanisms proposed to explain the observed trend of D-meson self-normalised yields at different relative multiplicities~\cite{Adam:2015ota}.

The measured near- and away-side peak yields and widths were compared, in the accessible kinematic ranges, with expectations from state-of-the-art models capable of producing peak observables, such as PYTHIA, POWHEG+PYTHIA, HERWIG, and EPOS. Among these models, PYTHIA8 and POWHEG+PYTHIA8 provide the best overall description of the data, HERWIG underestimates the near-side yields at low $\ptD$ and at high $\ptass$, while EPOS overestimates the near-side yields over the whole kinematic range, predicting also a steeper $\ptD$ dependence of the away-side yields.
A study of the influence of parton-level processes, as ISR, FSR, and MPI, on the shape of the final-state correlation distribution and on its peak properties was also performed with PYTHIA8 and POWHEG+PYTHIA8.
These studies are important not only for a better understanding of the underlying physics in pp collisions, but also for interpreting possible modifications of the correlation peaks in Pb--Pb collisions due to interactions of the heavy quarks in the quark--gluon plasma. This measurement is expected to become accessible during the LHC Run 3, with the upgraded ALICE detector~\cite{ALICE:2012dtf, ALICE:2013nwm}.


\newenvironment{acknowledgement}{\relax}{\relax}
\begin{acknowledgement}
\section*{Acknowledgements}

The ALICE Collaboration would like to thank all its engineers and technicians for their invaluable contributions to the construction of the experiment and the CERN accelerator teams for the outstanding performance of the LHC complex.
The ALICE Collaboration gratefully acknowledges the resources and support provided by all Grid centres and the Worldwide LHC Computing Grid (WLCG) collaboration.
The ALICE Collaboration acknowledges the following funding agencies for their support in building and running the ALICE detector:
A. I. Alikhanyan National Science Laboratory (Yerevan Physics Institute) Foundation (ANSL), State Committee of Science and World Federation of Scientists (WFS), Armenia;
Austrian Academy of Sciences, Austrian Science Fund (FWF): [M 2467-N36] and Nationalstiftung f\"{u}r Forschung, Technologie und Entwicklung, Austria;
Ministry of Communications and High Technologies, National Nuclear Research Center, Azerbaijan;
Conselho Nacional de Desenvolvimento Cient\'{\i}fico e Tecnol\'{o}gico (CNPq), Financiadora de Estudos e Projetos (Finep), Funda\c{c}\~{a}o de Amparo \`{a} Pesquisa do Estado de S\~{a}o Paulo (FAPESP) and Universidade Federal do Rio Grande do Sul (UFRGS), Brazil;
Ministry of Education of China (MOEC) , Ministry of Science \& Technology of China (MSTC) and National Natural Science Foundation of China (NSFC), China;
Ministry of Science and Education and Croatian Science Foundation, Croatia;
Centro de Aplicaciones Tecnol\'{o}gicas y Desarrollo Nuclear (CEADEN), Cubaenerg\'{\i}a, Cuba;
Ministry of Education, Youth and Sports of the Czech Republic, Czech Republic;
The Danish Council for Independent Research | Natural Sciences, the VILLUM FONDEN and Danish National Research Foundation (DNRF), Denmark;
Helsinki Institute of Physics (HIP), Finland;
Commissariat \`{a} l'Energie Atomique (CEA) and Institut National de Physique Nucl\'{e}aire et de Physique des Particules (IN2P3) and Centre National de la Recherche Scientifique (CNRS), France;
Bundesministerium f\"{u}r Bildung und Forschung (BMBF) and GSI Helmholtzzentrum f\"{u}r Schwerionenforschung GmbH, Germany;
General Secretariat for Research and Technology, Ministry of Education, Research and Religions, Greece;
National Research, Development and Innovation Office, Hungary;
Department of Atomic Energy Government of India (DAE), Department of Science and Technology, Government of India (DST), University Grants Commission, Government of India (UGC) and Council of Scientific and Industrial Research (CSIR), India;
Indonesian Institute of Science, Indonesia;
Istituto Nazionale di Fisica Nucleare (INFN), Italy;
Japanese Ministry of Education, Culture, Sports, Science and Technology (MEXT), Japan Society for the Promotion of Science (JSPS) KAKENHI and Japanese Ministry of Education, Culture, Sports, Science and Technology (MEXT)of Applied Science (IIST), Japan;
Consejo Nacional de Ciencia (CONACYT) y Tecnolog\'{i}a, through Fondo de Cooperaci\'{o}n Internacional en Ciencia y Tecnolog\'{i}a (FONCICYT) and Direcci\'{o}n General de Asuntos del Personal Academico (DGAPA), Mexico;
Nederlandse Organisatie voor Wetenschappelijk Onderzoek (NWO), Netherlands;
The Research Council of Norway, Norway;
Commission on Science and Technology for Sustainable Development in the South (COMSATS), Pakistan;
Pontificia Universidad Cat\'{o}lica del Per\'{u}, Peru;
Ministry of Education and Science, National Science Centre and WUT ID-UB, Poland;
Korea Institute of Science and Technology Information and National Research Foundation of Korea (NRF), Republic of Korea;
Ministry of Education and Scientific Research, Institute of Atomic Physics and Ministry of Research and Innovation and Institute of Atomic Physics, Romania;
Joint Institute for Nuclear Research (JINR), Ministry of Education and Science of the Russian Federation, National Research Centre Kurchatov Institute, Russian Science Foundation and Russian Foundation for Basic Research, Russia;
Ministry of Education, Science, Research and Sport of the Slovak Republic, Slovakia;
National Research Foundation of South Africa, South Africa;
Swedish Research Council (VR) and Knut \& Alice Wallenberg Foundation (KAW), Sweden;
European Organization for Nuclear Research, Switzerland;
Suranaree University of Technology (SUT), National Science and Technology Development Agency (NSDTA) and Office of the Higher Education Commission under NRU project of Thailand, Thailand;
Turkish Energy, Nuclear and Mineral Research Agency (TENMAK), Turkey;
National Academy of  Sciences of Ukraine, Ukraine;
Science and Technology Facilities Council (STFC), United Kingdom;
National Science Foundation of the United States of America (NSF) and United States Department of Energy, Office of Nuclear Physics (DOE NP), United States of America.
\end{acknowledgement}

\bibliographystyle{utphys}   
\bibliography{bibliography}

\newpage
\appendix

%
%

\section{The ALICE Collaboration}
\label{app:collab}
\small
\begin{flushleft}

S.~Acharya$^{\rm 143}$, 
D.~Adamov\'{a}$^{\rm 98}$, 
A.~Adler$^{\rm 76}$, 
J.~Adolfsson$^{\rm 83}$, 
G.~Aglieri Rinella$^{\rm 35}$, 
M.~Agnello$^{\rm 31}$, 
N.~Agrawal$^{\rm 55}$, 
Z.~Ahammed$^{\rm 143}$, 
S.~Ahmad$^{\rm 16}$, 
S.U.~Ahn$^{\rm 78}$
% ...

\section*{ Affiliation notes} 

$^{\rm I}$ Deceased\\
$^{\rm II}$ Also at: Italian National Agency for New Technologies, Energy and Sustainable Economic Development (ENEA), Bologna, Italy\\
$^{\rm III}$ Also at: Dipartimento DET del Politecnico di Torino, Turin, Italy\\
$^{\rm IV}$ Also at: M.V. Lomonosov Moscow State University, D.V. Skobeltsyn Institute of Nuclear, Physics, Moscow, Russia\\
$^{\rm V}$ Also at: Department of Applied Physics, Aligarh Muslim University, Aligarh, India
\\
$^{\rm VI}$ Also at: Institute of Theoretical Physics, University of Wroclaw, Poland\\
$^{\rm VII}$ Also at: University of Kansas, Lawrence, Kansas, United States\\

\section*{Collaboration Institutes}

$^{1}$ A.I. Alikhanyan National Science Laboratory (Yerevan Physics Institute) Foundation, Yerevan, Armenia\\
$^{2}$ AGH University of Science and Technology, Cracow, Poland\\
$^{3}$ Bogolyubov Institute for Theoretical Physics, National Academy of Sciences of Ukraine, Kiev, Ukraine\\
$^{4}$ Bose Institute, Department of Physics  and Centre for Astroparticle Physics and Space Science (CAPSS), Kolkata, India\\
$^{5}$ Budker Institute for Nuclear Physics, Novosibirsk, Russia\\
$^{6}$ California Polytechnic State University, San Luis Obispo, California, United States\\
$^{7}$ Central China Normal University, Wuhan, China\\
$^{8}$ Centro de Aplicaciones Tecnol\'{o}gicas y Desarrollo Nuclear (CEADEN), Havana, Cuba\\
$^{9}$ Centro de Investigaci\'{o}n y de Estudios Avanzados (CINVESTAV), Mexico City and M\'{e}rida, Mexico\\
$^{10}$ Chicago State University, Chicago, Illinois, United States\\
$^{11}$ China Institute of Atomic Energy, Beijing, China\\
$^{12}$ Chungbuk National University, Cheongju, Republic of Korea\\
$^{13}$ Comenius University Bratislava, Faculty of Mathematics, Physics and Informatics, Bratislava, Slovakia\\
$^{14}$ COMSATS University Islamabad, Islamabad, Pakistan\\
$^{15}$ Creighton University, Omaha, Nebraska, United States\\
$^{16}$ Department of Physics, Aligarh Muslim University, Aligarh, India\\
$^{17}$ Department of Physics, Pusan National University, Pusan, Republic of Korea\\
$^{18}$ Department of Physics, Sejong University, Seoul, Republic of Korea\\
$^{19}$ Department of Physics, University of California, Berkeley, California, United States\\
$^{20}$ Department of Physics, University of Oslo, Oslo, Norway\\
$^{21}$ Department of Physics and Technology, University of Bergen, Bergen, Norway\\
$^{22}$ Dipartimento di Fisica dell'Universit\`{a} 'La Sapienza' and Sezione INFN, Rome, Italy\\
$^{23}$ Dipartimento di Fisica dell'Universit\`{a} and Sezione INFN, Cagliari, Italy\\
$^{24}$ Dipartimento di Fisica dell'Universit\`{a} and Sezione INFN, Trieste, Italy\\
$^{25}$ Dipartimento di Fisica dell'Universit\`{a} and Sezione INFN, Turin, Italy\\
$^{26}$ Dipartimento di Fisica e Astronomia dell'Universit\`{a} and Sezione INFN, Bologna, Italy\\
$^{27}$ Dipartimento di Fisica e Astronomia dell'Universit\`{a} and Sezione INFN, Catania, Italy\\
$^{28}$ Dipartimento di Fisica e Astronomia dell'Universit\`{a} and Sezione INFN, Padova, Italy\\
$^{29}$ Dipartimento di Fisica e Nucleare e Teorica, Universit\`{a} di Pavia, Pavia, Italy\\
$^{30}$ Dipartimento di Fisica `E.R.~Caianiello' dell'Universit\`{a} and Gruppo Collegato INFN, Salerno, Italy\\
$^{31}$ Dipartimento DISAT del Politecnico and Sezione INFN, Turin, Italy\\
$^{32}$ Dipartimento di Scienze e Innovazione Tecnologica dell'Universit\`{a} del Piemonte Orientale and INFN Sezione di Torino, Alessandria, Italy\\
$^{33}$ Dipartimento di Scienze MIFT, Universit\`{a} di Messina, Messina, Italy\\
$^{34}$ Dipartimento Interateneo di Fisica `M.~Merlin' and Sezione INFN, Bari, Italy\\
$^{35}$ European Organization for Nuclear Research (CERN), Geneva, Switzerland\\
$^{36}$ Faculty of Electrical Engineering, Mechanical Engineering and Naval Architecture, University of Split, Split, Croatia\\
$^{37}$ Faculty of Engineering and Science, Western Norway University of Applied Sciences, Bergen, Norway\\
$^{38}$ Faculty of Nuclear Sciences and Physical Engineering, Czech Technical University in Prague, Prague, Czech Republic\\
$^{39}$ Faculty of Science, P.J.~\v{S}af\'{a}rik University, Ko\v{s}ice, Slovakia\\
$^{40}$ Frankfurt Institute for Advanced Studies, Johann Wolfgang Goethe-Universit\"{a}t Frankfurt, Frankfurt, Germany\\
$^{41}$ Fudan University, Shanghai, China\\
$^{42}$ Gangneung-Wonju National University, Gangneung, Republic of Korea\\
$^{43}$ Gauhati University, Department of Physics, Guwahati, India\\
$^{44}$ Helmholtz-Institut f\"{u}r Strahlen- und Kernphysik, Rheinische Friedrich-Wilhelms-Universit\"{a}t Bonn, Bonn, Germany\\
$^{45}$ Helsinki Institute of Physics (HIP), Helsinki, Finland\\
$^{46}$ High Energy Physics Group,  Universidad Aut\'{o}noma de Puebla, Puebla, Mexico\\
$^{47}$ Hiroshima University, Hiroshima, Japan\\
$^{48}$ Hochschule Worms, Zentrum  f\"{u}r Technologietransfer und Telekommunikation (ZTT), Worms, Germany\\
$^{49}$ Horia Hulubei National Institute of Physics and Nuclear Engineering, Bucharest, Romania\\
$^{50}$ Indian Institute of Technology Bombay (IIT), Mumbai, India\\
$^{51}$ Indian Institute of Technology Indore, Indore, India\\
$^{52}$ Indonesian Institute of Sciences, Jakarta, Indonesia\\
$^{53}$ INFN, Laboratori Nazionali di Frascati, Frascati, Italy\\
$^{54}$ INFN, Sezione di Bari, Bari, Italy\\
$^{55}$ INFN, Sezione di Bologna, Bologna, Italy\\
$^{56}$ INFN, Sezione di Cagliari, Cagliari, Italy\\
$^{57}$ INFN, Sezione di Catania, Catania, Italy\\
$^{58}$ INFN, Sezione di Padova, Padova, Italy\\
$^{59}$ INFN, Sezione di Pavia, Pavia, Italy\\
$^{60}$ INFN, Sezione di Roma, Rome, Italy\\
$^{61}$ INFN, Sezione di Torino, Turin, Italy\\
$^{62}$ INFN, Sezione di Trieste, Trieste, Italy\\
$^{63}$ Inha University, Incheon, Republic of Korea\\
$^{64}$ Institute for Gravitational and Subatomic Physics (GRASP), Utrecht University/Nikhef, Utrecht, Netherlands\\
$^{65}$ Institute for Nuclear Research, Academy of Sciences, Moscow, Russia\\
$^{66}$ Institute of Experimental Physics, Slovak Academy of Sciences, Ko\v{s}ice, Slovakia\\
$^{67}$ Institute of Physics, Homi Bhabha National Institute, Bhubaneswar, India\\
$^{68}$ Institute of Physics of the Czech Academy of Sciences, Prague, Czech Republic\\
$^{69}$ Institute of Space Science (ISS), Bucharest, Romania\\
$^{70}$ Institut f\"{u}r Kernphysik, Johann Wolfgang Goethe-Universit\"{a}t Frankfurt, Frankfurt, Germany\\
$^{71}$ Instituto de Ciencias Nucleares, Universidad Nacional Aut\'{o}noma de M\'{e}xico, Mexico City, Mexico\\
$^{72}$ Instituto de F\'{i}sica, Universidade Federal do Rio Grande do Sul (UFRGS), Porto Alegre, Brazil\\
$^{73}$ Instituto de F\'{\i}sica, Universidad Nacional Aut\'{o}noma de M\'{e}xico, Mexico City, Mexico\\
$^{74}$ iThemba LABS, National Research Foundation, Somerset West, South Africa\\
$^{75}$ Jeonbuk National University, Jeonju, Republic of Korea\\
$^{76}$ Johann-Wolfgang-Goethe Universit\"{a}t Frankfurt Institut f\"{u}r Informatik, Fachbereich Informatik und Mathematik, Frankfurt, Germany\\
$^{77}$ Joint Institute for Nuclear Research (JINR), Dubna, Russia\\
$^{78}$ Korea Institute of Science and Technology Information, Daejeon, Republic of Korea\\
$^{79}$ KTO Karatay University, Konya, Turkey\\
$^{80}$ Laboratoire de Physique des 2 Infinis, Ir\`{e}ne Joliot-Curie, Orsay, France\\
$^{81}$ Laboratoire de Physique Subatomique et de Cosmologie, Universit\'{e} Grenoble-Alpes, CNRS-IN2P3, Grenoble, France\\
$^{82}$ Lawrence Berkeley National Laboratory, Berkeley, California, United States\\
$^{83}$ Lund University Department of Physics, Division of Particle Physics, Lund, Sweden\\
$^{84}$ Moscow Institute for Physics and Technology, Moscow, Russia\\
$^{85}$ Nagasaki Institute of Applied Science, Nagasaki, Japan\\
$^{86}$ Nara Women{'}s University (NWU), Nara, Japan\\
$^{87}$ National and Kapodistrian University of Athens, School of Science, Department of Physics , Athens, Greece\\
$^{88}$ National Centre for Nuclear Research, Warsaw, Poland\\
$^{89}$ National Institute of Science Education and Research, Homi Bhabha National Institute, Jatni, India\\
$^{90}$ National Nuclear Research Center, Baku, Azerbaijan\\
$^{91}$ National Research Centre Kurchatov Institute, Moscow, Russia\\
$^{92}$ Niels Bohr Institute, University of Copenhagen, Copenhagen, Denmark\\
$^{93}$ Nikhef, National institute for subatomic physics, Amsterdam, Netherlands\\
$^{94}$ NRC Kurchatov Institute IHEP, Protvino, Russia\\
$^{95}$ NRC \guillemotleft Kurchatov\guillemotright  Institute - ITEP, Moscow, Russia\\
$^{96}$ NRNU Moscow Engineering Physics Institute, Moscow, Russia\\
$^{97}$ Nuclear Physics Group, STFC Daresbury Laboratory, Daresbury, United Kingdom\\
$^{98}$ Nuclear Physics Institute of the Czech Academy of Sciences, \v{R}e\v{z} u Prahy, Czech Republic\\
$^{99}$ Oak Ridge National Laboratory, Oak Ridge, Tennessee, United States\\
$^{100}$ Ohio State University, Columbus, Ohio, United States\\
$^{101}$ Petersburg Nuclear Physics Institute, Gatchina, Russia\\
$^{102}$ Physics department, Faculty of science, University of Zagreb, Zagreb, Croatia\\
$^{103}$ Physics Department, Panjab University, Chandigarh, India\\
$^{104}$ Physics Department, University of Jammu, Jammu, India\\
$^{105}$ Physics Department, University of Rajasthan, Jaipur, India\\
$^{106}$ Physikalisches Institut, Eberhard-Karls-Universit\"{a}t T\"{u}bingen, T\"{u}bingen, Germany\\
$^{107}$ Physikalisches Institut, Ruprecht-Karls-Universit\"{a}t Heidelberg, Heidelberg, Germany\\
$^{108}$ Physik Department, Technische Universit\"{a}t M\"{u}nchen, Munich, Germany\\
$^{109}$ Politecnico di Bari and Sezione INFN, Bari, Italy\\
$^{110}$ Research Division and ExtreMe Matter Institute EMMI, GSI Helmholtzzentrum f\"ur Schwerionenforschung GmbH, Darmstadt, Germany\\
$^{111}$ Russian Federal Nuclear Center (VNIIEF), Sarov, Russia\\
$^{112}$ Saha Institute of Nuclear Physics, Homi Bhabha National Institute, Kolkata, India\\
$^{113}$ School of Physics and Astronomy, University of Birmingham, Birmingham, United Kingdom\\
$^{114}$ Secci\'{o}n F\'{\i}sica, Departamento de Ciencias, Pontificia Universidad Cat\'{o}lica del Per\'{u}, Lima, Peru\\
$^{115}$ St. Petersburg State University, St. Petersburg, Russia\\
$^{116}$ Stefan Meyer Institut f\"{u}r Subatomare Physik (SMI), Vienna, Austria\\
$^{117}$ SUBATECH, IMT Atlantique, Universit\'{e} de Nantes, CNRS-IN2P3, Nantes, France\\
$^{118}$ Suranaree University of Technology, Nakhon Ratchasima, Thailand\\
$^{119}$ Technical University of Ko\v{s}ice, Ko\v{s}ice, Slovakia\\
$^{120}$ The Henryk Niewodniczanski Institute of Nuclear Physics, Polish Academy of Sciences, Cracow, Poland\\
$^{121}$ The University of Texas at Austin, Austin, Texas, United States\\
$^{122}$ Universidad Aut\'{o}noma de Sinaloa, Culiac\'{a}n, Mexico\\
$^{123}$ Universidade de S\~{a}o Paulo (USP), S\~{a}o Paulo, Brazil\\
$^{124}$ Universidade Estadual de Campinas (UNICAMP), Campinas, Brazil\\
$^{125}$ Universidade Federal do ABC, Santo Andre, Brazil\\
$^{126}$ University of Cape Town, Cape Town, South Africa\\
$^{127}$ University of Houston, Houston, Texas, United States\\
$^{128}$ University of Jyv\"{a}skyl\"{a}, Jyv\"{a}skyl\"{a}, Finland\\
$^{129}$ University of Kansas, Lawrence, Kansas, United States\\
$^{130}$ University of Liverpool, Liverpool, United Kingdom\\
$^{131}$ University of Science and Technology of China, Hefei, China\\
$^{132}$ University of South-Eastern Norway, Tonsberg, Norway\\
$^{133}$ University of Tennessee, Knoxville, Tennessee, United States\\
$^{134}$ University of the Witwatersrand, Johannesburg, South Africa\\
$^{135}$ University of Tokyo, Tokyo, Japan\\
$^{136}$ University of Tsukuba, Tsukuba, Japan\\
$^{137}$ Universit\'{e} Clermont Auvergne, CNRS/IN2P3, LPC, Clermont-Ferrand, France\\
$^{138}$ Universit\'{e} de Lyon, CNRS/IN2P3, Institut de Physique des 2 Infinis de Lyon , Lyon, France\\
$^{139}$ Universit\'{e} de Strasbourg, CNRS, IPHC UMR 7178, F-67000 Strasbourg, France, Strasbourg, France\\
$^{140}$ Universit\'{e} Paris-Saclay Centre d'Etudes de Saclay (CEA), IRFU, D\'{e}partment de Physique Nucl\'{e}aire (DPhN), Saclay, France\\
$^{141}$ Universit\`{a} degli Studi di Foggia, Foggia, Italy\\
$^{142}$ Universit\`{a} di Brescia, Brescia, Italy\\
$^{143}$ Variable Energy Cyclotron Centre, Homi Bhabha National Institute, Kolkata, India\\
$^{144}$ Warsaw University of Technology, Warsaw, Poland\\
$^{145}$ Wayne State University, Detroit, Michigan, United States\\
$^{146}$ Westf\"{a}lische Wilhelms-Universit\"{a}t M\"{u}nster, Institut f\"{u}r Kernphysik, M\"{u}nster, Germany\\
$^{147}$ Wigner Research Centre for Physics, Budapest, Hungary\\
$^{148}$ Yale University, New Haven, Connecticut, United States\\
$^{149}$ Yonsei University, Seoul, Republic of Korea\\

\end{flushleft} 
  
\end{document}